\documentclass[floatfix,groupedaddress,10pt,twocolumn]{revtex4-1}
\usepackage{graphics, color}
\usepackage{rotating} 
\usepackage{setspace} 
\usepackage{graphicx}
\usepackage[english]{babel}
\usepackage{amsmath,amsfonts,amssymb,verbatim,float,mathtools}
\usepackage{physics}
\usepackage[titletoc,title]{appendix} 
\usepackage{hyperref}
\hypersetup{%
	pdfborder = {0 0 0}}
\usepackage{cases}

\begin{document}
	
	\title{Decay rates and energies of free magnons and bound states in dissipative XXZ chains}
	\author{C. D. Parmee}
	\affiliation{T.C.M. Group, Cavendish Laboratory, University of Cambridge, JJ Thomson Avenue, Cambridge, CB3 0HE, U.K.}
	\author{N. R. Cooper}
	\affiliation{T.C.M. Group, Cavendish Laboratory, University of Cambridge, JJ Thomson Avenue, Cambridge, CB3 0HE, U.K.}
	\begin{abstract}
Chains of coupled two-level atoms behave as 1D quantum spin systems, exhibiting free magnons and magnon bound states. While these excitations are well studied for closed systems, little consideration has been given to how they are altered by the presence of an environment. This will be especially important in systems that exhibit nonlocal dissipation, e.g. systems in which the magnons decay due to optical emission. In this work, we consider free magnon excitations and two-magnon bound states in an XXZ chain with nonlocal dissipation. We prove that whilst the energy of the bound state can lie outside the two-magnon continuum of energies, the decay rate of the bound state has to always lie within the two-magnon continuum of decay rates. We then derive analytically the bound state solutions for a system with nearest-neighbour and next-nearest-neighbour XY interaction and nonlocal dissipation, finding that the inclusion of nonlocal dissipation allows more freedom in engineering the energy and decay rate dispersions for the bound states. Finally, we numerically study a model of an experimental set-up that should allow the realisation of dissipative bound states by using Rydberg-dressed atoms coupled to a photonic crystal waveguide (PCW). We demonstrate that this model can exhibit many key features of our simpler models.
	\end{abstract}
	\date{\today} 
	\maketitle

\section{Introduction}
One very interesting direction of recent research on ultracold atomic or molecular gases involves the study of the collective quantum dynamics of internal excitations of the atoms (or molecules) positioned in ordered arrays.  Such systems behave as strongly coupled two-level quantum systems (i.e. spin-1/2 systems), and can explore fundamental issues in the quantum dynamics of many-body systems subject to strong interparticle interactions \cite{Yan2013,Labuhn2016,Fukuhara2013}

A famous example of a strong-interaction phenomenon in quantum spins systems is provided by magnon bound states, first proposed by Bethe \cite{Bethe1931} more than 80 years ago. In this work, it was shown that magnon bound states could form in 1D spin-1/2 Heisenberg chain with nearest-neighbour interactions, lowering their energy compared to free magnons in the system. Subsequent work then extended this result to higher dimensions, anisotropic spin chains and arbitrary spin including solitons \cite{Wortis1963,Haldane1982,Southern1994,Schneider1981} and spin chains with long-range interactions \cite{Torrance1969,Majumdar1969,Ono1971,Letscher2018a}. Furthermore, magnon bound states have been studied in systems with frustration \cite{Kecke2007}, topological structure \cite{Qin2017,Qin2018} and in Floquet systems \cite{Agarwala2017,Kudo2009}. They have also recently been observed experimentally \cite{Fukuhara2013} and shown to have an important role in magnetisation switching \cite{Barker2013}, transport \cite{Krimphoff2017,Gana} and to have interesting effects on entanglement entropy \cite{Molter2014}. 

One key aspect in all of these studies is that the system is closed and so the question of bound state decay rates is not considered. However, if the system is coupled to an external environment, then the excitations will eventually decay and so it is natural to ask how long lived these excitations can be. For a system with local dissipation, the decay rate of both free excitations and bound states will be given by $m$ times the local decay rate \cite{Longo2014} where $m$ is the number of excitations. However, for systems involving radiative decay, the dissipation typically becomes nonlocal, where a range of decay rates to the environment exist, which are either superradiant (greater than the local decay rate) or subradiant (smaller than the local decay rate). In these scenarios, the relative decay rates of the free excitations and bound states becomes unclear. For example, is it possible for the decay rate of the bound states to be smaller than that of the free magnons?

In this work, we address the question of bound state decay rates in systems with nonlocal dissipation. We look at three models with a nearest-neighbour Ising interaction, which is crucial for the bound states to form, and different forms of XY interaction and nonlocal dissipation. The first two models are a nearest-neighbour and next-nearest-neighbour XY interaction for which we can obtain analytical results. The final model is an experimentally achievable setting in which to observe our results with Rydberg dressed atoms coupled to a photonic crystal waveguide.

The layout of the paper is as follows. In Sec \ref{sec:model}, we derive the general equations needed to obtain the energy and decay rate of the free excitations and bound states. In Sec \ref{sec:generaldecay}, we show that in general the decay rate of the bound state lies within the two-magnon decay rate continuum. Then in Sec \ref{sec:results}, we obtain the energies and decay rates for the three models. In Sec \ref{sec:discussion} we discuss our results and experimental implementation before drawing conclusions in Sec \ref{sec:conclusions}. 

\section{Model}
	\label{sec:model}
We consider a macroscopic number, $N$, of two-level systems fixed in position on a 1D optical lattice with spacing, $a$, and periodic boundary conditions. The atoms interact with an electromagnetic field which acts as an environment for the system. We assume the Markovian and Born approximations, which are valid provided the coupling between the system and environment is weak. These allow us to describe the system using a master equation approach. We will later discuss the validity of this approximation in relation to our results. The resultant master equation is given by
\begin{equation}\label{MasterEq}
\begin{split}	\dot{\hat{\rho}}(t)=\frac{i}{\hbar}\left[\hat{\rho}(t),\hat{H}\right]+\sum_{i,l}^{N}\frac{\Gamma_{il}}{2}\left(2\hat{\sigma}^-_i\hat{\rho}(t) \hat{\sigma}_l^+-\left\lbrace \hat{\sigma}_l^+\hat{\sigma}^-_i,\hat{\rho}(t)\right\rbrace\right),	\end{split}
\end{equation}
where the square brackets represent a commutator and curly brackets represent the anti-commutator. The spin operators are defined as $\hat{\sigma}^z_i=\ket{e_i}\bra{e_i}-\ket{g_i}\bra{g_i}$, $\hat{\sigma}^-_i=\ket{g_i}\bra{e_i}$ and $\hat{\sigma}^+_i=\ket{e_i}\bra{g_i}$, where $\ket{e_i}$ and $\ket{g_i}$ are the excited and ground states of the atom respectively. We  require that the eigenvalues of the matrix $\Gamma_{il}$ are all greater than or equal to zero, in order for Eq. \eqref{MasterEq} to describe decay of the excited state, driven by the operators $\hat{\sigma}^-_i$. Then the steady state density matrix is given by $\rho_{ss}=\ket{0}\bra{0}$ where $\ket{0}=\prod_{i}^{N}\ket{g_i}$. The Hamiltonian is given by
\begin{equation}\label{Hamiltonian}
\hat{H}=\hbar\Delta\sum_{i}^{N}\sigma^z_i+\sum_{i\neq l}^{N}\hbar V_{il}\hat{\sigma}_i^+\hat{\sigma}_l^-+\frac{\hbar J_z}{2}\sum_{i}^{N} \hat{\sigma}_i^z\hat{\sigma}_{i+1}^z.
\end{equation}
For the rest of the paper, we will work in units of $\hbar=1$. The Hamiltonian in Eq. \eqref{Hamiltonian} conserves the number of excitations in the system whilst the dissipator allows the excitations to decay. We can therefore talk about the dynamics of few-magnon excitations. To compute the energies and decay rates of one- and two-magnon excitations in our system, we employ a Green's function method. 

We first start with the single magnon Green's function, defined as $G(ij,t)=\Tr[\hat{\sigma}^-_i(t)\hat{\sigma}^+_j(0)\hat{\rho}(0)]\Theta(t)=\bra{0}\hat{\sigma}^-_i(t)\hat{\sigma}^+_j(0)\ket{0}\Theta(t)$, choosing the initial condition, $\rho(0)$, to be the pure state $\ket{0}\bra{0}$. The single magnon Green's function obeys the following equation
\begin{equation}\label{GreenfuncOneEOM}
\begin{split}
&\frac{dG(ij,t)}{dt}-\delta_{ij}\delta(t)=-i\Delta G(ij,t)-\frac{\Gamma}{2}G(ij,t)+\\
&4iJ_zG(ij,t)-i\sum_{p\neq j}^{N}\left(V_{pj}-i\frac{\Gamma_{pj}}{2}\right)G(ip,t),
\end{split}
\end{equation}
where $\Gamma=\Gamma_{ii}$. Fourier transforming Eq. \eqref{GreenfuncOneEOM} gives the spectrum of the single magnon states from the poles of	
\begin{equation}\label{SingleMagGreensFuncSoln}
G(k,\omega)=\lim\limits_{\epsilon\rightarrow0}i(\omega-E(k)+i\epsilon)^{-1},
\end{equation}
where 
\begin{equation}\label{SingleMagnonDispersion}
E(k)=-4J_z+\Delta-\frac{i\Gamma}{2}+\sum_{l=1}^{N}\left(2V_{l0}-i\Gamma_{l0}\right)\cos(kl)
\end{equation}
is the single magnon dispersion, with the real part corresponding to the energy and the magnitude of the imaginary part corresponding to the decay rate. 

For two magnons, we consider the Green's function $G(ij,lm;t)=\Tr[\hat{\sigma}^-_i(t)\hat{\sigma}^-_j(t)\hat{\sigma}^+_l\hat{\sigma}^+_m\rho(0)]\Theta(t)$, which obeys the following equation
\begin{equation}\label{GreenfuncTwoEOM}
\begin{split}
&\frac{dG(ij,lm;t)}{dt}-(1-\delta_{ij})\delta(t)(\delta_{il}\delta_{jm}+\delta_{im}\delta_{jl})=\\
&\left(-2i\Delta+8iJ_z-\Gamma-4iJ_z\delta_{m,l+1}\right)G(ij,lm;t)\\
&-i\sum_{p\neq l}^{N}J_{pl}G(ij,pm;t)-i\sum_{p\neq m}^{N}J_{pm}G(ij,pl;t)\\
&+2i\delta_{lm}\sum_{p\neq m}^{N}J_{pm}G(ij,pm;t),
\end{split}
\end{equation}
where $J_{pl}=V_{pl}-i\Gamma_{pl}/2$. This equation can be rewritten as a matrix equation and partially Fourier transformed with $
G(r,r',Q,\Omega)=\sum_{R}e^{-iRQ}\int_{-\infty}^{\infty}G(ij,lm,t)e^{i\Omega t}dt$
to give (see Appendix A)	
\begin{equation}\label{SolnGF2EOMFT}
\begin{split}
&G(r,r',Q,\Omega)=\Gamma(r,r',Q,\Omega)h(r)\\
&-\sum_{r''}^{N}K(r,r'',Q,\Omega)G(r'',r',Q,\Omega).
\end{split}
\end{equation}
where $r=r_i-r_j$, $r'=r_l-r_m$ and
\begin{equation}\label{FourierTransforms}
\begin{split}
&K(r,r';Q,\Omega)=\frac{2i}{N}\sum_{q\in BZ}\frac{\cos(qr')}{\Omega-S(q,Q)}\cross \\
&\left[4iJ_z\cos(q)-2i\left(V(r)-i\frac{\Gamma(r)}{2}\right)\cos(Qr/2)\right],\\
&\Gamma(r,r';Q,\Omega)=-\frac{2i}{N}\sum_{q\in BZ}\frac{\cos(qr')\cos(qr)}{\Omega-S(q,Q)}. 
\end{split}
\end{equation}
The momenta $q$ and $Q$ in Eq. \eqref{SolnGF2EOMFT} and Eq. \eqref{FourierTransforms} are the difference and sum of momenta, defined by $q=(k_1-k_2)/2$ and $Q=k_1+k_2$, where $k_1$ and $k_2$ are the momenta of the individual magnons. The momenta $q$ are summed over the Brillouin zone denoted by BZ. The function in the denominator of Eq. \eqref{FourierTransforms}, $S(q,Q)$, is the dispersion of two free magnons, given by
\begin{equation}\label{TwoMagnonDispersion}
\begin{split}
&S(q,Q)=E(Q/2+q)+E(Q/2-q)\\
&=-8J_z+2\Delta-i\Gamma+\sum_{j=1}^{N}\left(4V_{j0}-2i\Gamma_{j0}\right)\cos(jQa/2)\cos(jqa),
\end{split}
\end{equation}
which determines the poles of $\Gamma(r,r',Q,\Omega)$, whilst the two-magnon bound states are given by solutions to the determinant equation
\begin{equation}\label{DetEq}
\det\left[\delta(r,r'')+K(r,r'',Q,\Omega)\right]=0.
\end{equation}
Because of the nearest-neighbour Ising coupling, this determinant equation can be simplified to (see Appendix B)
\begin{equation}\label{NewBoundStates}
\begin{split}
&\left(1-\frac{1}{N}\sum_{q\in BZ}\frac{8J_z\cos^2(qa)}{\Omega-S(q,Q)}\right)\left(1+\frac{1}{N}\sum_{q'\in BZ}\frac{S(q',Q)-t}{\Omega-S(q',Q)}\right)\\
&+\frac{8J_z}{N^2}\sum_{q,q'\in BZ}\frac{\cos(qa)\cos(q'a)(S(q',Q)-t)}{\left[\Omega-S(q,Q)\right]\left[\Omega-S(q',Q)\right]}=0,
\end{split}
\end{equation}
where $t=8J_z-2\Delta+i\Gamma$.
In the limit $N\rightarrow \infty$, we can rewrite Eq. \eqref{NewBoundStates} as
\begin{equation}\label{Continuum}
\begin{split}
(\Omega+t)\left[\frac{I_0(\Omega,Q)}{8J_z}-I_0(\Omega,Q)I_2(\Omega,Q)+I_1(\Omega,Q)^2\right]=0,
\end{split}
\end{equation}
where 
\begin{equation}\label{Integrals}
I_m(\Omega,Q)=\int_{-\pi}^{\pi}\frac{\cos^m(q)}{\Omega-S(q,Q)}\frac{dq}{2\pi}.
\end{equation}
In Section IV, we shall find the energies and decay rates of the bound states by solving Eq. \eqref{Continuum} (or Eq. \eqref{NewBoundStates} where appropriate) for three specific forms of the XY interaction and nonlocal dissipation: a nearest-neighbour model, next-nearest-neighbour model and a photonic crystal waveguide model. Note that $\Omega=t=-8J_z+2\Delta-i\Gamma$ is always a solution to Eq. \eqref{Continuum}. However, this solution always lies within the two-magnon energy continuum. In general, we will dismiss any solutions that lie inside the two-magnon energy continuum where the bound state is no longer well defined because it can scatter into the continuum states and become a resonance. While it is possible to have bound states that exist in the scattering continuum \cite{Hsu2016}, these usually occur when the system has certain symmetries that protect the state, which we are not aware of existing in our models. 

\section{General Decay Rates of Bound States}
\label{sec:generaldecay}
We first show that in general, for any model with nonlocal dissipation of the form given in the master equation, Eq. \eqref{MasterEq}, the decay rate of the bound state always lies within the two-magnon decay rate continuum, i.e. the bound state cannot decay more quickly or slowly than its constituent parts. To show this, we consider Eq. \eqref{MasterEq} rewritten in diagonal form
\begin{equation}\label{MasterEqDiagonalForm}
\begin{split}	\dot{\hat{\rho}}(t)=i\left[\hat{\rho}(t),\hat{H}\right]+\sum_{k}\left(2\hat{J}^-_k\hat{\rho}(t) \hat{J}_k^+-\left\lbrace \hat{J}_k^+\hat{J}^-_k,\hat{\rho}(t)\right\rbrace\right).
\end{split}
\end{equation}
Here, $\hat{J}_k$ is a decay operator for mode $k$, given by $\hat{J}_k^-=\sqrt{\gamma_k}\sum_{i}^{N}c_i^k\hat{\sigma}^-_i$, where $c_i^k$ is the $i^{th}$ component of the $k^{th}$ eigenvector of $\Gamma_{il}/2$ and $\gamma_k$ is the corresponding eigenvalue. For a periodic or large enough system, the eigenvector components are given by $c_i^k=e^{ikr_i}/\sqrt{N}$. To determine the decay rate of the bound state, we focus on the initial dynamics of the pure state density matrix, $\hat{\rho}(0)=\ket{Q}\bra{Q}$ where $\ket{Q}$ is the wavefunction of a bound state with momentum $Q$, given by
\begin{equation}
\begin{split}	
&\ket{Q}=\sum_{ij}^{N}\alpha_Qf_Q(|r_i-r_j|)e^{iQ(r_i+r_j)/2}\hat{\sigma}_i^+\hat{\sigma}_j^+\ket{0},
\end{split}
\end{equation}
where $f_Q(r)$ is some localised function that determines the spatial decay of the bound state, with $r=|r_i-r_j|$, and $\alpha_Q$ is a normalisation constant given by $\alpha_Q=1/\left(2N\sum_{r\neq 0}|f_Q(r)|^2\right)$. The equation of motion for a pure bound state density matrix at short initial times is given by
\begin{equation}\label{BSEOM}
\begin{split}	
&\frac{d\rho_Q(t)}{dt}\approx-\sum_{k}8\gamma_k|\alpha_QF(Q/2-k)|^2\rho_Q(t),
\end{split}
\end{equation}
where $\rho_Q(t)=\bra{Q}\hat{\rho}(t)\ket{Q}$ and
\begin{equation}
F(Q/2-k)=\sum_{r\neq 0}f_Q(r)e^{ir(Q/2-k)}
\end{equation}is the Fourier transform of the localised function. At later times, there can be the population of coherences between the bound state and scattering states, which we have neglected. We can see that the bound state density matrix has a decay rate of $4\tilde{\gamma}_Q$, where $\tilde{\gamma}_Q\equiv\sum_{k}^{}2\gamma_k|\alpha_QF(Q/2-k)|^2$, which is weighted sum of all single magnon decay rates. Note that $\tilde{\gamma}_Q$, is equivalent to the decay rate we will obtain from our Green's function method. 
	
For local dissipation where $\gamma_k=\gamma\equiv\Gamma/2$, the sum over $k$ in $\tilde{\gamma}_Q$ can be completed to give 
\begin{equation}\label{id}
\sum_k2|\alpha_QF(Q/2-k)|^2=1,
\end{equation}
and so the decay rate of the bound state wavefunction (which is half the decay rate of the pure density matrix) is $2\gamma$ as expected. For nonlocal dissipation, in order to have a bound state decay rate that exists below the two-magnon decay rate continuum, we would need 
\begin{equation}
\tilde{\gamma}_Q=\sum_k2\gamma_k|\alpha_QF(Q/2-k)|^2<\gamma_{\text{min}},
\end{equation}
where $\gamma_{\text{min}}$ is the smallest decay rate for a single magnon. However, using Eq. \eqref{id}, we can rewrite this condition as
\begin{equation}
\sum_k2(\gamma_k-\gamma_{\text{min}})|\alpha_QF(Q/2-k)|^2<0.
\end{equation}
Both $|\alpha_QF(Q/2-k)|^2$ and $\gamma_k-\gamma_{\text{min}}$ are always positive, which means this condition can never be fulfilled. The lowest decay rate that could possibly be achieved for the bound state is the lowest decay rate that can be achieved for two free magnons, although this may not always obey the bound state equation. The same argument applies for showing that the bound state cannot have a decay rate above the two-magnon decay rate continuum, such that
\begin{equation}
\sum_k2(\gamma_k-\gamma_{\text{max}})|\alpha_QF(Q/2-k)|^2>0,
\end{equation}
where $\gamma_{\text{max}}$ is the largest decay rate in the system. Again $|\alpha_QF(Q/2-k)|^2>0$, but $\gamma_k-\gamma_{\text{max}}<0$, so this condition can not be satisfied and the bound state decay rate must always lie within the two-magnon decay rate continuum.

\section{Results}
\label{sec:results}
\subsection{Nearest-Neighbour Model}
Having now shown in general that the decay rate of the bound state always lies within the two-magnon decay rate continuum, we now look at three specific models for dissipative bound states. The first model we consider is where all interactions are nearest-neighbour (NN). The energies and decay rates of the one and two free magnon states are given by
\begin{equation}\label{NN1-2magnons}
\begin{split}
&\Re[E(k)]=-4J_z+\Delta+2V_{12}\cos(ka)\\
&|\Im[E(k)]|=\frac{\Gamma}{2}+\Gamma_{12}\cos(ka)\\
&\Re[S(q,Q)]=-8J_z+2\Delta+4V_{12}\cos(Qa/2)\cos(qa)\\
&|\Im[S(q,Q)]|=\Gamma+2\Gamma_{12}\cos(Qa/2)\cos(qa).\\
\end{split}
\end{equation}
Solving Eq. \eqref{Continuum} gives the following bound state solution (see Appendix C)
\begin{equation}\label{NNSolution}
\begin{split}
\Omega(Q)=-4J_z+2\Delta-i\Gamma+\frac{(2V_{12}-i\Gamma_{12})^2}{4J_z}\cos^2(Qa/2),
\end{split}
\end{equation}
which can be written in terms of the energy and decay rate as
\begin{equation}\label{NN Solution}
\begin{split}
&\Re[\Omega(Q)]=-4J_z+2\Delta+\frac{4V_{12}^2-\Gamma_{12}^2}{4J_z}\cos^2(Qa/2)\\
&|\Im[\Omega(Q)]|=\Gamma+\frac{V_{12}\Gamma_{12}}{J_z}\cos^2(Qa/2).\\
\end{split}
\end{equation}
These expressions first appeared in Ref. \cite{Longo2014c}, although we analyse them in more detail here. For the expressions in Eqs. \eqref{NN Solution}, there are limits to the parameters we can choose for the solutions to satisfy the bound state equation, Eq. \eqref{Continuum}. However, provided we choose $V_{12}$ and $\Gamma_{12}$ such that the energy term in Eq. \eqref{NN Solution} lies below the two-magnon energy continuum, then we find the bound state equation is always satisfied. We also have to impose $\Gamma/2\geq|\Gamma_{12}|$ in order for the dissipator to always give decay.

Comparing the bound state solution Eq. \eqref{NN Solution} to the free magnon dispersions in Eq. \eqref{NN1-2magnons}, we see the energy and decay rate of the bound state depend on a mixture of the interaction and dissipation. The presence of nonlocal dissipation creates a negative shift in energy compared to the XY interaction, which means that the bound state energy is shifted further from the two-magnon energy continuum than in a closed system. This is important as the effects of nonlocal dissipation will not only cause the bound state to decay, but will alter its dynamics travelling through the lattice meaning that even if the bound state has a very small decay rate, it is not sufficient to ignore environmental effects. Furthermore, due to nonlocal dissipation, there is more freedom to engineer the bound state energy and decay than in a closed system. For example, the bound state energy band can be made entirely flat by choosing $V_{12}=\Gamma_{12}/2$. Also, by choosing $V_{12}=0$ such that there is no XY interaction, the bound state experiences only local dissipation, with a decay rate of $\Gamma$, whereas the one and two free magnons still experience nonlocal dissipation. Finally, in the limit where $V_{12},\Gamma_{12}\ll J_z$, the effects of the XY interaction and nonlocal dissipation become negligible, with the energy of the bound state tending to $-4J_z$ and the decay rate tending to $\Gamma$ which would be expected for an Ising model with local dissipation. 

\begin{figure}
	\hspace*{-0.5cm}
	\includegraphics[scale=0.5,clip,angle=0]{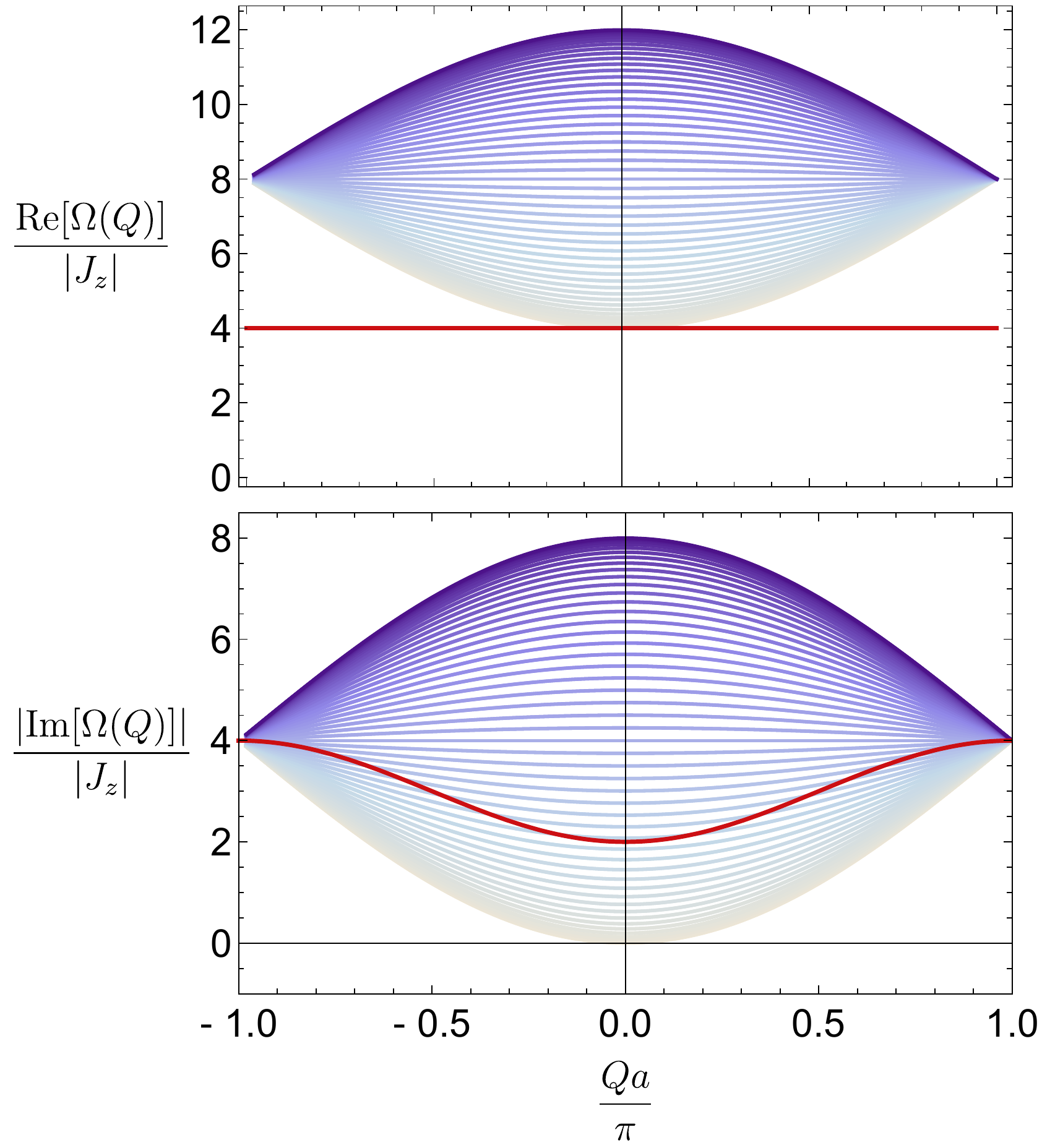}
	\caption{Energy (top) and decay rate (bottom) of the bound state for a NN system for $J_z<0$ and $\Gamma=2|\Gamma_{12}|$. The bound state solution is shown in red whilst the shaded region represents the two-magnon continuum. The parameters used are $\Gamma_{12}/|J_z|=-2$ and $V_{12}/|J_z|=-1$ which give the smallest possible decay rate for the bound state while keeping the energy separate from the continuum.}
	\vspace{0cm}
	\label{NNminlifetimes}
\end{figure}

The relative signs of the XY interaction, nonlocal dissipation and Ising interaction allow the bound state decay rate to be tuned such that it is either entirely subradiant or superradiant, with the most super- or subradiant decay at $Qa=0$ and a decay rate of $\Gamma$ at the band edge, $Qa=\pm \pi$. To find how subradiant or superradiant it is possible to make the bound state, we extremise the decay rate of the bound state with respect to the parameters $V_{12}$ and $\Gamma_{12}$, while still obeying the constraint that the bound state energy must lie below the two-magnon energy continuum. We also maintain a fixed decay rate $\Gamma$ (otherwise there is always a trivial minimal decay rate with $\Gamma=\Gamma_{12}=0$). We find the extremal decay rates and corresponding energies are given by 
\begin{equation}\label{MaxMinLifetime}
\begin{split}
&\Re[\Omega(Q)]=-4J_z+2\Delta\\
&|\Im[\Omega(Q)]|=\Gamma\mp 2J_z\cos^2(Qa/2),\\
\end{split}
\end{equation}
where the negative sign gives the maximal (minimal) decay rate and the positive sign gives the minimal (maximal) decay rate for $J_z<0$ ($J_z>0$). The largest values for $\Gamma_{12}$ and $V_{12}$ occur when the bound state makes contact with the energy continuum at $Qa=0$. In figure \ref{NNminlifetimes}, we show the minimal decay rate solution for $J_z<0$ and $\Gamma=2|\Gamma_{12}|$.
The bound state decay rate lies in the two-magnon decay rate continuum as expected and is smaller than half the free magnon decay rates at $Qa=\pi$ and $2/3$ of the continuum at $Qa=0$, with the lowest energy bands of the two-magnon continuum having the smallest decay rates. For the maximal decay rate solution, the results are similar to Figure \ref{NNminlifetimes}, but the decay rates reverse, with the lowest energy bands having the highest decay rates and the bound state solution having a larger decay rate than most of the two-magnon decay rate continuum.
 
\subsection{Next-Nearest-Neighbour Model}
The NN model studied in the previous section demonstrated many features of dissipative bound states, but also missed some qualitative features of bound states with longer range hopping. We therefore consider a next-nearest-neighbour (NNN) model, finding that the inclusion of additional site interactions produces important differences in the properties of the bound state compared to a NN model. The one and two free magnon energies and decay rates are given by
\begin{equation}\label{NNN1-2magnons}
\begin{split}
\Re[E(k)]=&-4J_z+\Delta+2V_{12}\cos(ka)+2V_{13}\cos(2ka)\\
|\Im[E(k)]|=&\frac{\Gamma}{2}+\Gamma_{12}\cos(ka)+\Gamma_{13}\cos(2ka)\\
\Re[S(q,Q)]=&-8J_z+2\Delta+4V_{12}\cos(Qa/2)\cos(qa)\\
&+4V_{13}\cos(Qa)\cos(2qa)\\
|\Im[S(q,Q)]|=&\Gamma+2\Gamma_{12}\cos(Qa/2)\cos(qa)\\
&+2\Gamma_{13}\cos(Qa)\cos(2qa).\\
\end{split}
\end{equation}
The bound state solution is given by (see Appendix D)
\begin{equation}\label{NNNBSFull}
\begin{split}
&\Omega(Q)=-8J_z+2\Delta-i\Gamma+4J_{13}\cos(Qa)+\frac{J_{12}^2\cos^2(Qa/2)}{J_z}\\
&+\frac{J_{12}^2\cos^2(Qa/2)J_{13}\cos(Qa)}{2J_z^2}+\frac{8J_z^2}{2J_z+J_{13}\cos(Qa)},
\end{split}
\end{equation} 
where $J_{12}=V_{12}-i\Gamma_{12}/2$ and $J_{13}=V_{13}-i\Gamma_{13}/2$. Writing in terms of the energy and decay rate gives
\begin{equation}\label{NNNAnalytics}
\begin{split}
&\Re[\Omega(Q)]=-8J_z+2\Delta+\frac{4V_{12}^2-\Gamma_{12}^2}{4J_z}\cos^2(Qa/2)\\
&+\frac{V_{13}(4V_{12}^2-\Gamma_{12}^2)-2\Gamma_{13}\Gamma_{12}V_{12}}{8J_z^2}\cos(Qa)\cos^2(Qa/2)\\
&+4V_{13}\cos(Qa)+\frac{16J_z^2(4J_z+2V_{13}\cos(Qa))}{(4J_z+2V_{13}\cos(Qa))^2+(\Gamma_{13}\cos(Qa))^2}\\
&|\Im[\Omega(Q)]|=\Gamma+\frac{V_{12}\Gamma_{12}}{J_z}\cos^2(Qa/2)+2\Gamma_{13}\cos(Qa)\\
&+\frac{\Gamma_{13}(4V_{12}^2-\Gamma_{12}^2)+8V_{13}\Gamma_{12}V_{12}}{16  J_z^2}\cos(Qa)\cos^2(Qa/2)\\
&-\frac{16J_z^2\Gamma_{13}\cos(Qa)}{(4J_z+2V_{13}\cos(Qa))^2+(\Gamma_{13}\cos(Qa))^2}.
\end{split}
\end{equation}
As for the NN model, there is a constraint on the values of the dissipative couplings to ensure the magnons always decay, which is  $\Gamma/2\geq|\Gamma_{12}+\Gamma_{13}|$. Likewise, we have to choose parameters that satisfy the bound state condition Eq. \eqref{Continuum}, finding again that provided the energy of the bound state lies below the continuum, then Eq. \eqref{Continuum} is satisfied. Our NNN bound state solution is the same as that found in Ref. \cite{Letscher2018a} but with a complex XY interaction. This is also true of our NN result in Eq. \eqref{NN Solution}, which can be obtained by taking the bound state result in Ref. \cite{Wortis1963} with a complex XY interaction.

The inclusion of an additional site in the XY interaction and nonlocal dissipation results in a more complex bound state solution than in the NN model. Looking at the terms in Eq. \eqref{NNNAnalytics} in more detail, we see that the NN solution in Eq. \eqref{NN Solution} can be recovered by letting $V_{13},\Gamma_{13}=0$, and that now we have additional terms due to two-site hopping processes and a term that mixes the NN and NNN parameters. Because of the new magnon hopping terms, the decay rate of the bound state is no longer fixed to be $\Gamma$ at $Qa=\pm \pi$ as was the case for NN interactions, and the smallest and largest decay rates do not have to occur at $Qa=0$ anymore. Therefore the inclusion of NNN interactions allows more freedom in choosing at what momenta $Q$ the bound state can have its highest or smallest decay rate. However, we can now no longer engineer an entirely flat energy band due to the presence of both $\cos(Qa/2)$ and $\cos(Qa)$ terms (unless trivially the NNN couplings are set to zero). Looking at the limit of $V_{13},\Gamma_{13},V_{12},\Gamma_{12}\ll J_z$, we find Eq. \eqref{NNNAnalytics} simplifies to
\begin{equation}\label{NNNweakXY}
\begin{split}
\Re[\Omega(Q)]&\approx-4J_z+2\Delta+2V_{13}\cos(Qa)\\
|\Im[\Omega(Q)]|&\approx\Gamma+\Gamma_{13}\cos(Qa),\\
\end{split}
\end{equation}
We find that there is now always a contribution to the decay rate from the NNN interactions, that means even tightly confined bound states still experience the effects of nonlocal dissipation, which was not the case for the NN model. We can also see that the smallest decay rate will occur at $Qa=0$ ($Qa=\pm\pi$) and largest decay rate at $Qa=\pm\pi$ ($Qa=0$) for $\Gamma_{13}<0$ ($\Gamma_{13}>0$). 

\begin{figure}
	\hspace*{-0.5cm}
	\includegraphics[scale=0.5,clip,angle=0]{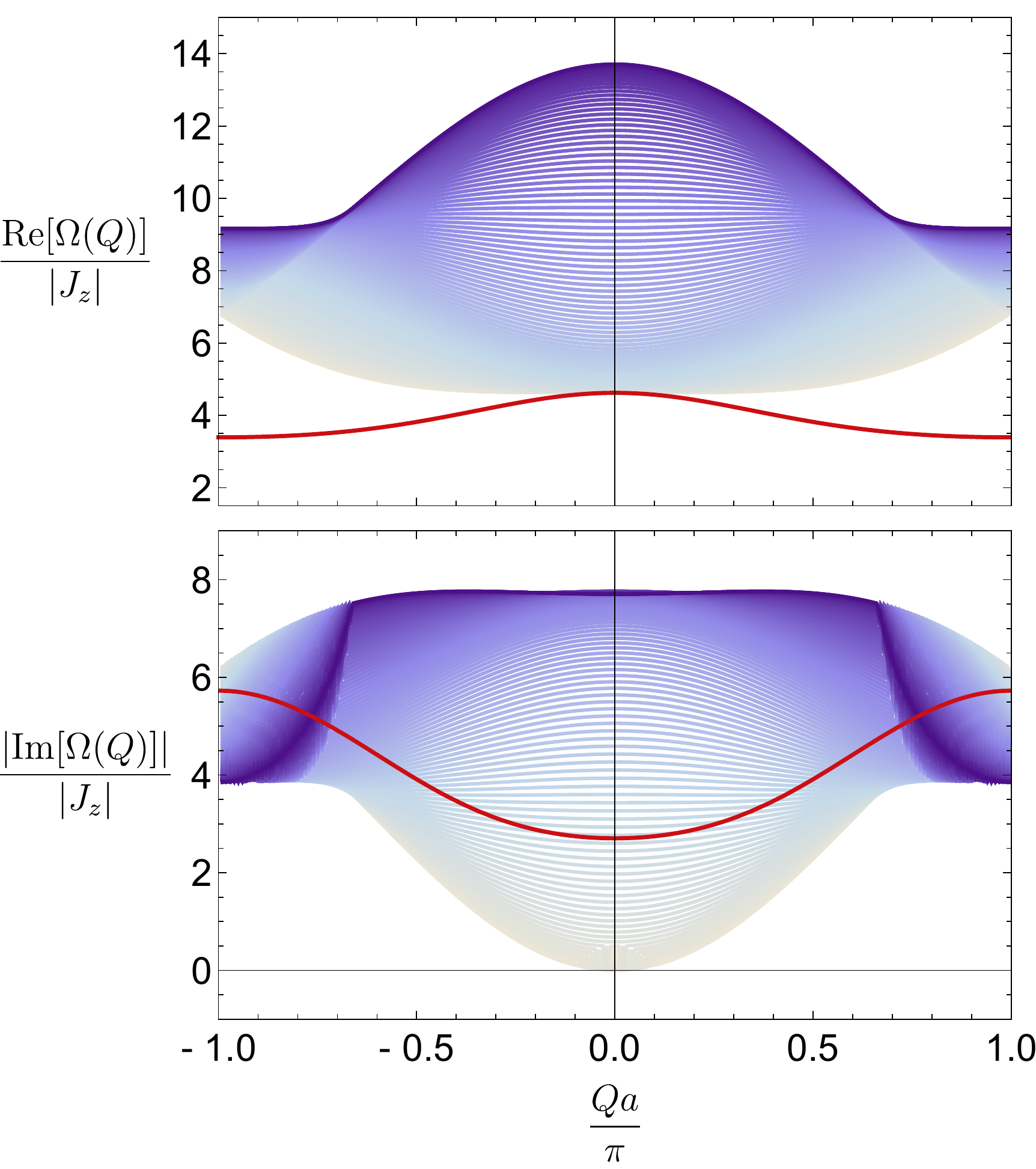}
	\caption{Energy (top) and decay rate (bottom) of the bound state for a NNN system with $\Gamma=2|(\Gamma_{12}+\Gamma_{13})|$ and $J_z<0$. The bound state solution is shown in red whilst the shaded region represents the two-magnon continuum. The parameters used are $V_{12}/|J_z|=1.135$, $V_{13}/|J_z|=0.293$, $\Gamma_{12}/|J_z|=1.926$ and $\Gamma_{13}/|J_z|=-0.578$, which give the smallest possible decay rate for the bound state while keeping the energy separate from the continuum.}
	\vspace{-0cm}
	\label{NNNminlifetimes}
\end{figure}

We now extremise the NNN bound state decay rate for a fixed $\Gamma$ with respect to the parameters $V_{12}$, $V_{13}$, $\Gamma_{12}$ and $\Gamma_{13}$ to find the smallest and largest decay rates the bound state can have while its energy remains separate from the two-magnon energy continuum. Due to the complexity of Eqs. \eqref{NNNAnalytics}, we solve this numerically, finding that the solution with minimal (maximal) decay rate occurs when
$V_{12}=\pm1.135 J_z$, $V_{13}=-0.293 J_z$, $\Gamma_{12}=\pm 1.926 J_z$  and 
$\Gamma_{13}=0.578 J_z$, and the maximal (minimal) solution occurs when $V_{12}=\mp1.135 J_z$, $V_{13}=-0.293 J_z$, $\Gamma_{12}=\pm 1.926 J_z$ and 
$\Gamma_{13}=-0.578 J_z$ for $J_z<0$ ($J_z>0$), where in both cases, we are free to choose the positive or negative sign. The largest values of all parameters occur when the bound state energy makes contact with the two-magnon energy continuum at $Qa=0$ as was the case for the NN interactions. In figure \ref{NNNminlifetimes}, we show the minimal solution with $J_z<0$ and $\Gamma=2|(\Gamma_{12}+\Gamma_{13})|$.
Again, we find the decay rate of the bound state lies within the two-magnon decay rate continuum, with the bound state having a smaller decay rate than $30\%$ of the continuum at $Qa=\pi$ and up to $70\%$ of the continuum at $Qa=0$. We should note there is a second minimal (maximal) decay rate solution with parameters $V_{12}=\Gamma_{12}=0$, $\Gamma_{13}=+0.402 J_z$ and $V_{13}=-0.827 J_z$ and maximal (minimal) solution for $V_{12}=\Gamma_{12}=0$, $\Gamma_{13}=-0.402 J_z$ and $V_{13}=-0.827 J_z$ for $J_z<0$ ($J_z>0$). However, we have not shown this solution as it is more unphysical due to the absence of the NN terms.

\subsection{Photonic Crystal Waveguide Model}
We now study one final model which should be an experimentally realisable set-up to study dissipative bound states. We consider Rydberg dressed two-level atoms that are coupled to a photonic crystal waveguide (PCW). Systems of two-level atoms where one state is a Rydberg state or Rydberg dressed are already well studied as realisable quantum simulators \cite{Schauss2018,Weimer2010,Whitlock2016,Nguyen2018,Glaetzle2015}. Likewise, PCWs are also gaining attention as a method for quantum simulation and quantum information processing due to the high tunability of the interactions between coupled quantum emitters \cite{Hartmann2016,Hood2016,Goban2014,Douglas2015,Gonz}. For atoms coupled to a PCW, photons emitted from the atoms can propagate to other atoms along the chain, which mediates an effective XY interaction and nonlocal dissipation. For a single mode in a dissipative PCW, the XY interaction and nonlocal dissipation are given by \cite{Calajo2016} $V_{il}=\Im{A_{ij}}$ and $\Gamma_{il}=2\Re{A_{ij}}$, where $A_{ij}$ is of the form
\begin{equation}\label{PCWInteraction}
A_{ij}=\frac{J_{xy}e^{iK|r_{ij}|}}{2\sqrt{1-(\delta/(2J)+i\gamma_c/(4J))^2}}.
\end{equation}
The parameter $J_{xy}$ is the coupling of the atoms to the PCW, $J$ is an energy scale determining the PCW bandwidth, and $K_{wq}a=\pi-\arccos{(\delta/(2J)+i\gamma_c/(4J))}=k_{wg}a+i\kappa_{wg}a$ is the PCW wavevector. The PCW wavevector depends on the detuning, $\delta=(\omega_{eg}-\omega_{wg})$, of the atomic transition frequency, $\omega_{eg}$, from the photon mode frequency, $\omega_{wg}$, and also the loss rate of photons from the PCW, $\gamma_c$. If $|\delta/J|<2$, then the photon lies within the bandwidth and can propagate along the PCW with a group velocity given by the denominator of Eq. \eqref{PCWInteraction}, $2\sqrt{1-(\delta/(2J)+i\gamma_c/(4J))^2}$. However, if $|\delta/J|>2$, then the photon cannot propagate and instead exponentially decays along the PCW.

In order for bound states to form, we also need an Ising interaction. This can be engineered by dressing \cite{Glaetzle2015} either the excited state, $\ket{e}$ or ground state, $\ket{g}$, of an atom with a Rydberg state $\ket{r}$, giving a new state $\ket{\tilde{e}}=\ket{e}+\beta\ket{r}$ where $\beta=\Omega_d/2\Delta_d$, set by the drive $\Omega_d$ and detuning $\Delta_d$ that couple $\ket{e}$ to $\ket{r}$. The atoms then interact with an Ising interaction of the form 
\begin{equation}
U_{il}=\frac{U_0}{1+(|r_i-r_l|/R_c)^6},
\end{equation}
where $U_0=\hbar\Omega_d^4/8\Delta_d^3$ and $R_c$ is some cut off length to the interaction. For small $R_c$, this is a good approximation to a NN Ising interaction. The sign and magnitude of $U_0$ can be fixed by the laser detuning and it is also possible to add additional XY interactions between the atoms which gives more freedom in tuning $V_{ij}$ separately from $\Gamma_{ij}$. 
	
For the PCW system, the one and two free magnon energies and decay rates are given by
\begin{equation}\label{PCWMagnonDispersion}
\begin{split}
\Re[E(k)]&=-4J_z+\Delta+f(k)+f(-k)\\
|\Im[E(k)]|&=\frac{\Gamma}{2}+g(k)+g(-k)\\
\Re[S(q,Q)]&=-8J_z+2\Delta+f(Q/2+q)+f(Q/2-q)\\
&+f(-Q/2+q)+f(-Q/2-q)\\
|\Im[S(q,Q)]|&=\Gamma+g(Q/2+q)+g(Q/2-q)\\
&+g(-Q/2+q)+g(-Q/2-q),
\end{split}
\end{equation}
where $\Delta=V_{11}/2+\delta/2+\delta_{\text{add}}$, with $\delta_{\text{add}}$ being an additional detuning to those from the waveguide, and
\small
\begin{equation}\label{PCW1-2magnons}
\begin{split}
&f(k)=\left(\frac{\Gamma\sin((k_{wg}+k)a)+V_{11}[\cos((k_{wg}+k)a)-e^{-\kappa_{wg}a}]}{e^{\kappa_{wg}a}+e^{-\kappa_{wg}a}-2\cos((k_{wg}+k)a)}\right).\\
&g(k)=\left(\frac{\Gamma[\cos((k_{wg}+k)a)-e^{-\kappa_{wg}a}]-V_{11}\sin((k_{wg}+k)a)}{e^{\kappa_{wg}a}+e^{-\kappa_{wq}a}-2\cos((k_{wg}+k)a)}\right)\\
\end{split}
\end{equation}
\normalsize
For the rest of this section, we will choose the additional detuning, $\delta_{\text{add}}$ such that $\Delta=0$ and so we can ignore the contributions to energy from the onsite term, $V_{11}$ and detuning from the waveguide mode $\delta$. We will also work with $J_z<0$.
\begin{figure}
	\hspace*{0cm}
	\includegraphics[scale=0.6,clip,angle=0]{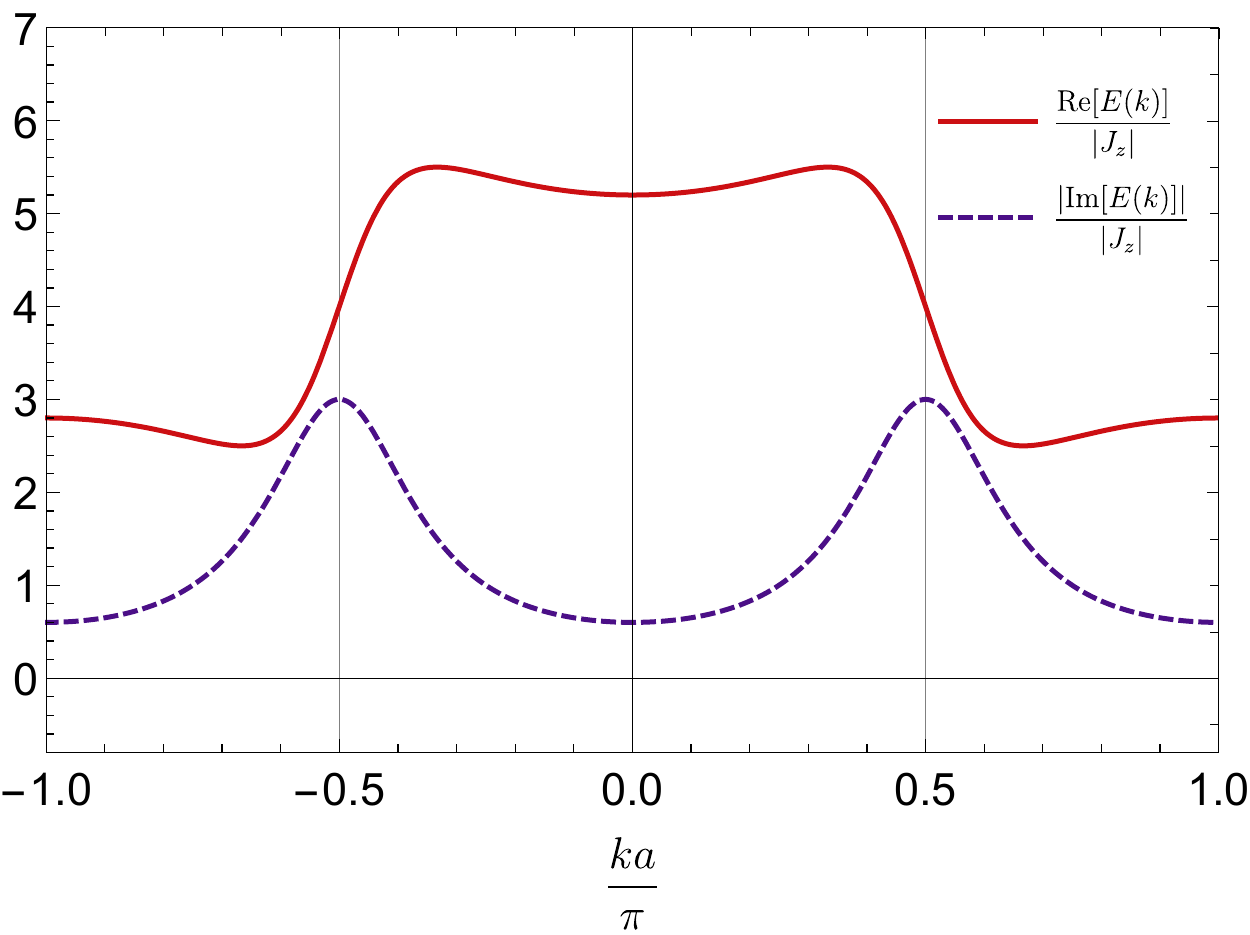}
	\caption{Energy and decay rate of a single magnon for the PCW system with $\gamma_c/J=2$, $\delta/J=0$ and $J_{xy}/|J_{z}|=3$. The energy is shown by the red (solid) line and the decay rate by the purple (dashed) line. The largest decay rates occur when $k=\pm k_{wg}$, shown by the grey lines.}
	\vspace{0cm}
	\label{SingleMagnonDispersionPlot}
\end{figure}

In figure \ref{SingleMagnonDispersionPlot}, we plot the energy and decay rate of the single magnon dispersion for $\gamma_c/J=2$, $\delta/J=0$ and $J_{xy}/|J_{z}|=3$. If $|\delta/J|<2$ and $\gamma_c/J$ is small, then about the points $k=\pm k_{wg}$, the decay rate is well modelled by two Lorentzians with a width of $4\sinh(\kappa_{wg}a/2)$ and maximum value of $\Gamma/[4\tanh(\kappa_{wg}a/2)]$. Similarly, the energy of the magnon is well described by the derivative of a Lorentzian with width $4\sinh(\kappa_{wg}a/2)$ and maximal (minimal) values given by $\pm\Gamma/[8\sinh(\kappa_{wg}a/2)]$. As $\gamma_c/J$ decreases (and so $\kappa_{wg}\rightarrow0$), the energies of the magnons and decay rates about $k=\pm k_{wg}$ diverge within the photonic bandwidth ($|\delta/J|<2$). However, outside the bandwidth ($|\delta/J|>2$), the energy of the magnon is bounded and its decay rate drops to zero as $\gamma_c\rightarrow0$, leaving the system effectively closed. The single magnon dispersions can be thought of as the hybridisation of a photon propagating through the waveguide with a dispersion $\omega_k=\omega_{wg}-J\cos(k)$ and momentum $k$, and a single atom with energy $\omega_{eg}$. 

\begin{figure*}	
	\hspace*{-0.5cm}
	\includegraphics[scale=1.2,clip,angle=0]{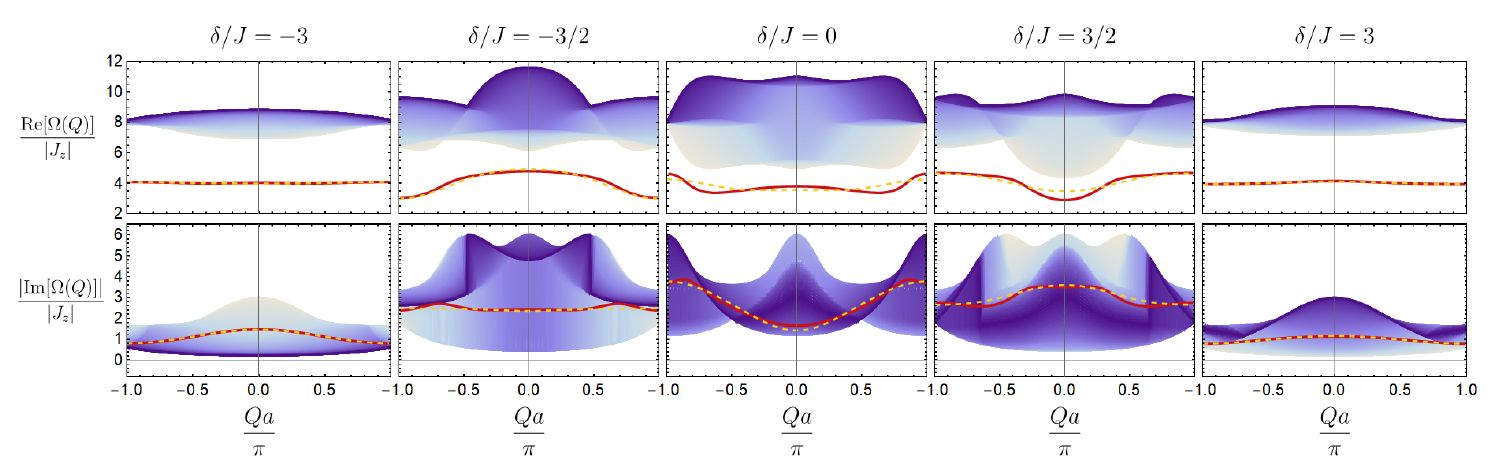}
	\caption{Examples of the two-magnon bound states that can form in the PCW model for a system size of $N=99$ with parameters $\gamma_c/J=2$, $J_{xy}/|J_z|=3$ and $\delta/J=(-3,-1.5,0,1.5,3)$. The top panels show the energy of the bound state and the lower panels show the decay rate. The red line represents the bound state solution and the shaded region represents the continuum of two-magnon states. We find that the bound state energy lies below the two-magnon energy continuum and the decay rate of the bound state always lies within the two-magnon decay rate continuum. When $\delta/J<-2$, the bound state decay rate is always lower than that of the lowest energy bands whilst if $\delta/J>2$, then the decay rate of the bound state is larger than the lowest energy bands. Note that for $\delta/J=0$, the decay rates of the lowest energy bands are obscured by the highest energy bands as they share the same decay rate. We also show the NNN bound state result from Eq. \eqref{NNNAnalytics} with the dashed orange line. We see the NNN result agree well with the waveguide results when $\kappa_{wg}a$ is large.}
	\label{BoundStatess}
	\vspace{-0cm}
\end{figure*}

We now look at the bound state solutions in the PCW and discuss their properties. The bound state condition, Eq. \eqref{Continuum}, is too complex to be solved analytically, so we instead tackle the problem numerically for finite sized systems by solving Eq. \eqref{NewBoundStates}. In Figure \ref{BoundStatess}, we plot some typical solutions of Eq. \eqref{NewBoundStates} for a system size of $N=99$, with $\gamma_c/J=2$, $J_{xy}/|J_z|=3$ and for $\delta/J=(-3,-1.5,0,1.5,3)$.
We see that bound state decay rate lies within the two-magnon decay rate continuum as expected, and is smaller than the decay rate of the lowest energy bands of the continuum for $\delta/J<-2$, but larger than the decay rate of the lowest energy bands of the continuum for $\delta/J>2$. For intermediate detunings, whether the bound state decay rate is smaller or larger than the lowest energy bands depends on the momentum of the bound state. As for the NNN model, we find the minimal and maximal decay rate of the bound state is no longer constrained to occur at $Qa=0$ and that the decay rate at $Qa=\pi$ is not given by $\Gamma$ as a consequence of the long-range interactions. If $\kappa_{wg}$ is large enough, then the bound state solutions are well modelled by the NNN analytics due to the exponential decay of the PCW interaction. This can be seen by the close agreement between the NNN and PCW bound state solutions when $\delta/J=\pm3$, which gives the largest $\kappa_{wg}$. For intermediate detunings, the agreement is not as good, but can be made increasingly better for larger $\gamma_c/J$.

\begin{figure}
	\hspace*{-0cm}
	\includegraphics[scale=0.35,clip,angle=0]{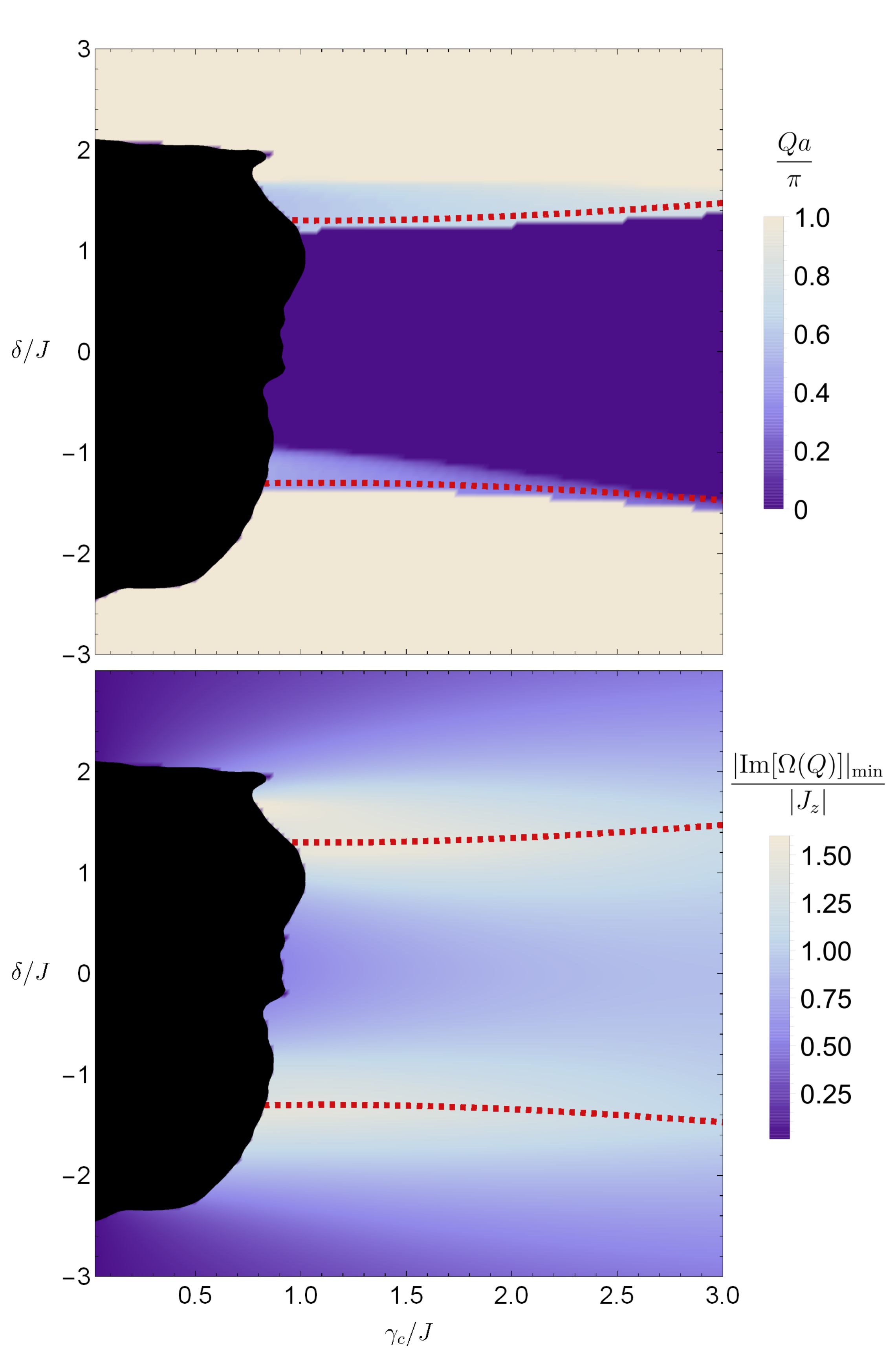}
	\caption{(a) Momentum at which the smallest decay rate of the bound state occurs for a system size of $N=99$ with $J_{xy}/|J_z|=1.5$. We see there is a clear transition between the smallest decay rate occurring at $Qa=\pi$ when $|\delta/J|\gtrsim1.4$, and $Qa=0$ for $|\delta/J|\lesssim1.4$. The red dashed lines show when $\Gamma_{13}$ changes sign which explains the transition as described in the main text. The black region shows where the bound state solution starts to merge with the two-magnon continuum. (b) Magnitude of the smallest decay rate. We see that when the cross over in momentum occurs when $\Gamma_{13}=0$, the decay rate increases, but decreases again as $\Gamma_{13}$ becomes larger.}
	\label{Qplotlifetimes}
	\vspace{-0cm}
\end{figure}

In figure \ref{Qplotlifetimes}, we plot the momentum for which the bound state has the smallest decay rate as a function of $\delta/J$ and $\gamma_c/J$. We find that there is a transition between the bound state having the smallest decay rate at $Qa=0$ when $|\delta/J|<1.4$ to $Qa=\pi$ when $|\delta/J|>1.4$. This transition can be explained by looking at the weak XY limit of the NNN bound state solutions given by Eq. \eqref{NNNweakXY}. In the weak limit, we find that the momentum where the decay rate of the bound state is smallest transitions from $Qa=0$ to $Qa=\pi$ when $\Gamma_{13}$ changes sign. We show when $\Gamma_{13}=0$ in figure \ref{Qplotlifetimes} by the red dashed lines, and find it agrees well with the transition in the PCW, with $\Gamma_{13}<0$ when $|\delta/J|>1.4$. The transition moves to larger values of $|\delta/J|$ as $\gamma_c/J$ increases, and also becomes sharper as the NNN solution becomes a better approximation to the PCW results.

\begin{figure}
	\hspace*{-0.5cm}
	\includegraphics[scale=0.6,clip,angle=0]{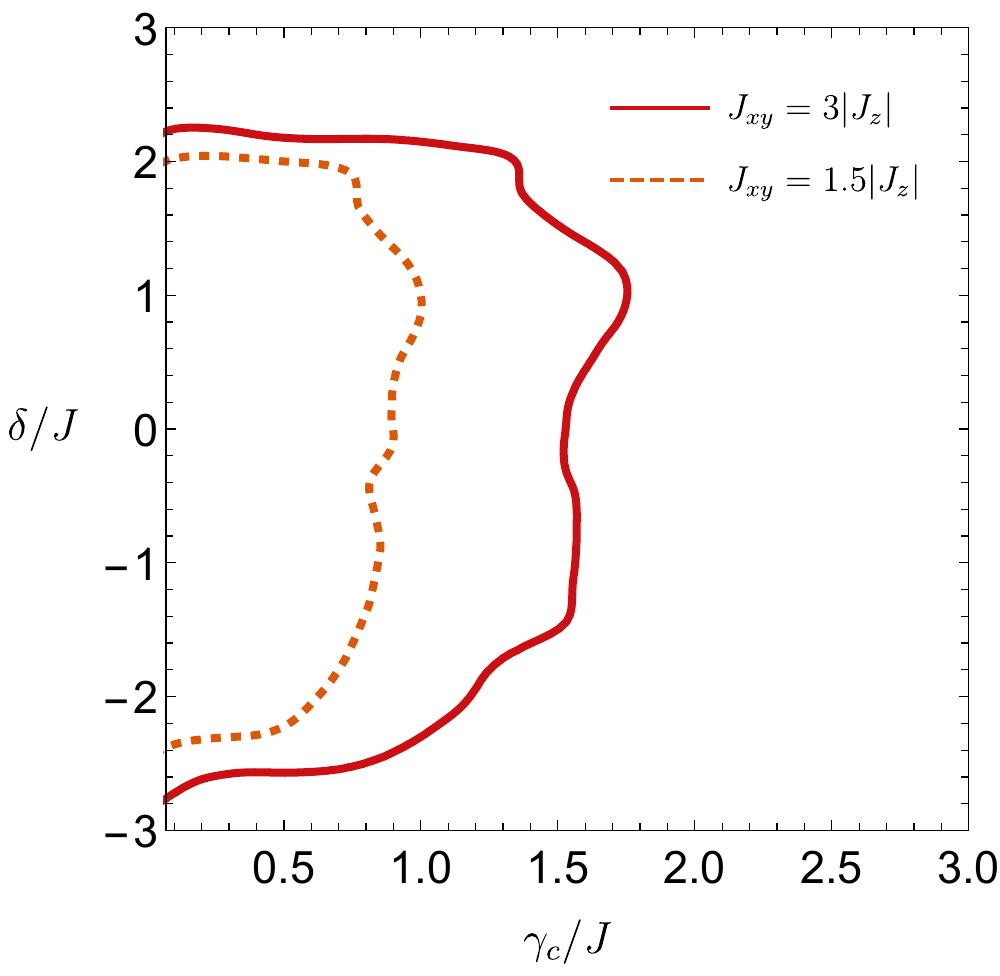}
	\caption{Diagram of when the bound state can form for the PCW model for a system size of $N=99$ with $J_{xy}/|J_z|=1.5$ (dashed line) and $J_{xy}/|J_z|=3$ (solid line). Between the $\delta$ axis and the bound state line, the bound state energy starts to join the two-magnon energy continuum for some or all momenta, $Q$. Outside this region, the bound state energy lies separate from the two-magnon energy continuum for all momenta $Q$. We see that the bound state can not remain separate from the two-magnon energy continuum at low $\gamma_c/J$ near the band edge or inside the bandwidth, but can remain separate from the two-magnon energy continuum everywhere else.}
	\label{BSFormationPlot}
	\vspace{-0cm}
\end{figure}

Finally, we discuss how the bound state formation depends on $\delta/J$ and $\gamma_c/J$. Figure \ref{BSFormationPlot} shows where the bound state rejoins the two magnon energy continuum as a function of $\delta/J$ and $\gamma_c/J$.
We find there is a region inside the bandwidth that extends along the $\gamma_c/J$ axis where the bound state joins the continuum and that, as $J_{xy}/|J_z|$ increases, this region also increases in size. The reason the bound state starts to rejoin the continuum for small $\gamma_c/J$ inside the bandwith is due to the diverging strength of the single magnon energy around $k=\pm k_{wg}$. For increasingly large systems, more momentum modes around these points are allowed and so the energy range of the two magnon continuum grows until the bound state is absorbed. However, outside the bandwidth and in the small $\gamma_c/J$ limit, the bound state energy can remain separate from the two-magnon energy continuum for any value of $J_{xy}/|J_z|$ provided $\delta/J$ is large enough. This is because the two-magnon energy continuum is now bounded as $\gamma_c/J\rightarrow0$ and so bound states can remain separate from the continuum. As mentioned in our discussion of the single magnon dispersion, the imaginary part of the PCW interaction, Eq. \eqref{PCWInteraction}, becomes negligible in this limit, and so the system becomes closed, with the decay rate of the bound state dropping to zero. When $\gamma_c/J$ becomes large, or when $|\delta/J|\gg2$, the XY interaction becomes increasingly shorter ranged due to the exponential decay, until eventually it is negligible compared to the Ising interaction. In this limit, the bound state is well separated from the two-magnon energy continuum with the bound state energy tending to $-4J_z$ and the decay rate tending to $\Gamma$.

Our analysis of a PCW has shown how many features of dissipative bound states can be obtained for a single photonic mode and how, for large $\kappa_{wg}a$, the PCW is well described by the NNN analytics. For a single mode, it is not possible to obtain the NN results, no matter how large $\kappa_{wg}a$ is. To see why this is the case, we look at the NNN bound state solution in Eq. \eqref{NNNBSFull}. We can see that for an exponentially decaying function, $J_{13}\sim J_{12}^2/J_z$, which means that there is always a NNN contribution to the bound state solution that is of the order of the NN parts, so the NNN contribution cannot be ignored. However, it could be possible to engineer more exotic XY interactions by combining many modes or coupling to more than one waveguide. This could also be done in parallel with different Rydberg dressing schemes or allowing other interactions, such as dipole interactions, to occur between atoms.

\section{Discussion}
\label{sec:discussion}
We have shown that two-magnon bound states can generally form in dissipative spin chains with XY and Ising interactions. We find the inclusion of nonlocal dissipation not only gives the bound state a momentum dependent decay rate, but also alters the bound state energy compared to a closed system or system with local dissipation. Nonlocal dissipation also allows for a greater degree of freedom in engineering the energy and decay rate of the bound state. We have shown that the decay rate of the bound state cannot be smaller or larger than its constituent free magnons. Nevertheless, it is still possible to achieve bound states that have a decay rate much lower than a large proportion of the two-magnon decay rate continuum. 

We now discuss the experimental set-up of the PCW in more detail. To engineer the bound states, we need to choose an appropriate scheme for Rydberg dressing for the atoms. Rydberg dressing has already been achieved experimentally \cite{Zeiher2016} with $\prescript{87}{}{\text{Rb}}$ atoms, taking the Rubidium hyperfine states $\ket{g}=\ket{1,-1}$ and $\ket{e}=\ket{2,-2}$ and dressing with a suitable Rydberg state of $\ket{r}=\prescript{31}{}{P_{1/2}}$. Therefore, it should be possible to engineer suitable Ising-like interactions with NN or even beyond NN range. The PCW itself can be realised with a SiO alligator waveguide \cite{Goban2014,Douglas2015} with high tunability over the allowed modes and loss processes. Previous experiments with cold atoms in waveguides have used Caesium, but it should be possible to engineer a waveguide suitable for Rubidium \cite{Perrella2018}. When studying the bound states, one has to be careful not to violate the Markovian approximation. For the Markovian approximation to be valid, it is required that the time for a photon to travel down the PCW is negligible compared to the decay rate of the atoms \cite{Calajo2016}. This gives the condition
\begin{equation}\label{markov}
\frac{\sqrt{J_{xy}/J}}{2\sqrt{1-(\delta/(2J)+i\gamma_c/(4J)^2)}}\ll\frac{1}{\sqrt{(N-1)a}},
\end{equation}
which is satisfied provided the coupling of the atoms to the waveguide, $J_{xy}$, is weak and also that the detuning is away from the band edge at $\delta=\pm2J$ when $\gamma_c/J$ is small. The expression Eq. \eqref{markov} also shows that the system needs to be finite to not violate the Markovian approximation. However, we have checked and found that there are bound state solutions with similar properties to those in the main text for finite size systems with open boundary conditions. Therefore, it should be possible to observe many of our bound states results for large enough finite sized systems with open boundary conditions or periodic boundary conditions.

Finally, measurement of the bound state decay rate and energy should be possible by observing the emission when the bound state decays. Following the steps outlined in Ref. \cite{Longo2014c}, the emission properties of the bound state are given by the correlator $g(t,\boldsymbol{r})=\langle\boldsymbol{\hat{E}}^{(-)}(t,\boldsymbol{r})\boldsymbol{\hat{E}}^{(+)}(t,\boldsymbol{r})\rangle$ which can be calculated from the electric field, $\boldsymbol{\hat{E}}^{(-)}(t,\boldsymbol{r})$. For decay of a pure bound state, $\hat{\rho}(0)=\ket{Q}\bra{Q}$, the correlator $g(t,\boldsymbol{r})$ is given by
\begin{equation}
\begin{split}
&\frac{g(t,\boldsymbol{r})}{|\eta W(\boldsymbol{r})|^2}=\sum_{k}4|\alpha_Q F(Q/2-k)|^2\biggr[\delta_{Q-k,\Delta^Q_k\sin(\beta)/c}e^{-4\tilde{\gamma}_Qt_{r}}\\
&+\frac{\gamma_{k+Q}}{\tilde{\gamma}_Q-\gamma_k}\delta_{k,\Delta^k_0\sin(\beta)/c}\left(e^{-2\gamma_kt_{r}}-e^{-4\tilde{\gamma}_Qt_{r}}\right)\biggr],
\end{split}
\end{equation}
where $t_{r}\equiv t-r/c$, $\Delta_0^k=\text{Re}[E(k)]$, $\Delta_k^Q=\text{Re}[\Omega(Q)]-\text{Re}[E(k)]$, $\eta=\omega_{eg}^2/(4\pi\epsilon_0c)$ and $W(\boldsymbol{r})=\boldsymbol{d}/r-\boldsymbol{r}(\boldsymbol{d}.\boldsymbol{r})/r^3$ is the far-field dipole emission profile. There are two contributions to the emission of the bound state; one from the decay of the bound state to a single magnon with momentum $k$, and one from the decay of a single magnon to the ground state. The delta functions determine the emission angle $\beta$ for each of these decay processes in terms of the momentum and energy of the bound state and single magnons, where $\beta$ is defined from the perpendicular axis from the spin chain. The total emission is then a sum over all these processes. The quantity $|\alpha_Q F(Q/2-k)|^2$ that determined the decay rate of the bound state also plays a crucial role in the angular dependence of the emission, which was noted in \cite{Longo2014c}. By examining the spatial and temporal emission of the bound state, it should be possible to determine its energy and decay rate for a given momentum $Q$.

In future work, it would be interesting to extend our results to $m$ magnon-bound states and to see how the decay rates of different magnon sectors compare to one another. Given our proof that the two-magnon bound state decay rate must lie within the continuum of decay rates, it seems likely that this would also be true for $m$ magnon states, and possibly also true for magnon states with larger spin and in systems of higher dimension. It would also be interesting to study different forms of dissipators and find systems where the bound state can have a decay rate that lies outside the two-magnon continuum.

\section{Conclusions}
\label{sec:conclusions}
We have studied the energies and decay rates of one and two free magnons and two-magnon bound states in an XXZ model with nonlocal dissipation.  We have proved that in general the decay rate of the bound state must lie within the decay rate continuum of two free magnons. We have then examined three examples of dissipative bound states in more detail, first looking at two forms of the XY interaction analytically; a nearest-neighbour model and next-nearest-neighbour model. We have found that the inclusion of nonlocal dissipation leads to momentum dependent decay rates and changes in the energy of the bound state compared to a closed system or a system with local dissipation. The nonlocal dissipation also allows a higher degree of tunability in the energies and decay rates of the bound states. Finally, in our third example, we have numerically studied an experimentally realisable model to observe dissipative bound states using Rydberg dressed atoms coupled to a photonic crystal waveguide, which demonstrates many key features of our simpler models and can also be used to obtain our  next-nearest-neighbour results within certain parameter regimes.

\section{Acknowledgements}
This work was supported by EPSRC Grant Nos. EP/K030094/1 and EP/P009565/1 and by the Simons Foundation. Statement of compliance with EPSRC policy framework on research data: All data accompanying this publication are directly available within the publication.

\begin{appendices}

\section{Deriving the Bound State Determinant Equation}
Below, we outline the steps to obtain the bound state equation in Eq. \eqref{NewBoundStates}.
For an open quantum system, provided the Liouvillian operator is time independent, any Heisenberg operator will obey the adjoint master equation, given by \cite{Breuer2007}
\begin{equation}\label{AdjointMasterEq}
\frac{d\hat{A}(t)}{dt}=i[\hat{H},\hat{A}(t)]+\sum_{i,l}^{N}\frac{\Gamma_{il}}{2}\left(2\hat{\sigma}_l^+\hat{A}(t) \hat{\sigma}_i^--\left\lbrace \hat{\sigma}_l^+\hat{\sigma}^-_i,\hat{A}(t) \right\rbrace \right).
\end{equation}
Therefore, the Green's function $\Tr(\hat{A}(t)\hat{B(0)}\hat{\rho}(0))=\bra{0}\hat{A}(t)\hat{B}\ket{0}$, with the initial condition $\hat{\rho}(0)=\ket{0}\bra{0}$, will obey
\begin{equation}\label{GreenfuncABsimplified}
\begin{split}
&\bra{0}\frac{d\hat{A}(t)}{dt}\hat{B}\ket{0}=i\bra{0}\hat{A}(t)[\hat{B},\hat{H}]\ket{0}\\
&-\sum_{i,l}^{N}\frac{\Gamma_{il}}{2}\bra{0}\hat{A}(t) \hat{\sigma}_l^+\hat{\sigma}^-_i\hat{B}\ket{0}.
\end{split}
\end{equation} 
For the two-magnon Green's function, $G(ij,lm;t)=\bra{0}\hat{\sigma}^-_i(t)\hat{\sigma}^-_j(t)\hat{\sigma}^+_l\hat{\sigma}^+_m\ket{0}\Theta(t)$, this gives
\begin{equation}\label{GreenfuncTwoEomAPP}
\begin{split}
&\frac{dG(ij,lm;t)}{dt}-(1-\delta_{ij})\delta(t)(\delta_{il}\delta_{jm}+\delta_{im}\delta_{jl})=\\
&\left(-2i\Delta+4i\sum_{a\neq m}^{N}U_{am}+4i\sum_{a\neq l}^{N}U_{al}-\Gamma-8iU_{lm}\right)G(ij,lm;t)\\
&-i\sum_{p\neq l}^{N}J_{pl}G(ij,pm;t)-i\sum_{p\neq m}^{N}J_{pm}G(ij,pl;t)\\
&+2i\delta_{lm}\sum_{p\neq m}^{N}J_{pm}G(ij,pm;t),
\end{split}
\end{equation}
where $J_{pl}=V_{pl}-i\Gamma_{pl}/2$.
In order to solve Eq. \eqref{GreenfuncTwoEomAPP}, it will be useful to view it as a matrix equation \cite{Majlis2000} given by $(\boldsymbol{\mathcal{L}}+\boldsymbol{\delta\mathcal{L}})\boldsymbol{G}=\boldsymbol{\mu}\boldsymbol{h}$,
where the matrices are defined as 
\begin{equation}\label{MatrixDefns}
\begin{split}
&\mathcal{L}(lm,pv,t-t')=i\delta(t-t')\delta_{vm}J_{pl}+i\delta(t-t')\delta_{vl}J_{pm}+\\
&\delta(t-t')\delta_{pl}\delta_{vm}\left(\frac{d}{dt'}+2i\Delta-4i\sum_{a\neq m}^{N}U_{am}-4i\sum_{a\neq l}^{N}U_{al}+\Gamma\right),\\
&\mathcal{\delta L}(lm,pv,t-t')=-i\delta(t-t')\delta_{pm}\delta_{lm}J_{vl}\\
&-i\delta(t-t')\delta_{vl}\delta_{lm}J_{pl}+8i\delta(t-t')\delta_{pl}\delta_{vm}U_{pv},\\
&h(ij,pv)=\delta_{ip}\delta_{jv}(1-\delta_{ij}),\\
&\mathcal{\mu}(lm,pv,t-t')=\delta(t-t')(\delta_{pl}\delta_{vm}+\delta_{pm}\delta_{vl}),\\
\end{split}
\end{equation}To solve Eq. \eqref{GreenfuncTwoEomAPP}, we now follow the same steps taken by Wortis \cite{Wortis1963} by introducing the function $\Gamma(ij,lm;t)=G(il;t)G(jm;t)+G(im;t)G(jl;t)$, where $G(jl;t)$ is the single magnon Green's function. We find that $\Gamma(ij,lm;t)$ obeys Eq. \eqref{GreenfuncTwoEomAPP} without the last two terms and no $1-\delta_{ij}$ term. Viewed in terms of matrices, this means $\boldsymbol{\mathcal{L}}\boldsymbol{\Gamma}=\boldsymbol{\mu}$ and so we can write $\boldsymbol{\mathcal{L}}=\boldsymbol{\mu}\boldsymbol{\Gamma}^{-1}$. This allows Eq. \eqref{GreenfuncTwoEomAPP} to be rewritten as
\begin{equation}\label{NewMatrixEquation}
\begin{split}
&\Gamma(ij,ab,t)h(ij)-G(ij,ab,t)\\
&=\int_{-\infty}^{\infty}\sum_{pv}^{N}\sum_{lm}^{N}\Gamma(lm,ab,t)\delta\mathcal{L}(lm,pv,t-t')G(ij,pv,t')\\
&=\int_{-\infty}^{\infty}\sum_{pv}^{N}K(ab,pv,t-t')G(ij,pv,t'),\\
\end{split}
\end{equation}
where in the last line we have defined
\begin{equation}
\begin{split}
&K(ab,pv;t)=8iU_{pq}\Gamma(pv,ab,t)\\
&-i(J_{pv}/2)\left(\Gamma(vv,ab,t)+\Gamma(pp,ab,t)\right).
\end{split}
\end{equation}
In order to obtain the bound state solutions, we now need to partially Fourier transform Eq. \eqref{NewMatrixEquation}. The Fourier transform of $\Gamma(ij,ab;t)$ is given by
\begin{equation}\label{FourierTransformsGamma1}
	\Gamma(ij,ab;\Omega)=\int_{-\infty}^{\infty}\Gamma(ij,ab,t)e^{i\Omega t}dt.
\end{equation}
By using the definition of $\Gamma(ij,ab;t)$ and the Fourier transform of the single magnon Green's function, this can be written as
\begin{equation}\label{FourierTransformsGamma2}
\begin{split}
&\Gamma(ij,ab;\Omega)=\sum_{k_1\in BZ}\sum_{k_2\in BZ}\left(\frac{e^{ik_1r_{ia}+ik_2r_{jb}}+e^{ik_1r_{ib}+ik_2r_{ja}}}{N}\right)\cross\\
&\int_{-\infty}^{\infty}\int_{-\infty}^{\infty}\int_{-\infty}^{\infty}\tilde{G}(k_1,\omega_1)\tilde{G}(k_2,\omega_2)e^{i(\Omega-\omega_1-\omega_2) t}dt \frac{d\omega_1}{2\pi} \frac{d\omega_2}{2\pi},
\end{split}
\end{equation}
where $r_{ia}=r_i-r_a$. We now rewrite the momentum sums using the sum and difference of momenta, $Q=k_1+k_2$ and $q=(k_1-k_2)/2$, and also the sum and difference of coordinates $R=(r_i+r_j)/2$, $r=r_i-r_j$ and $R'=(r_a+r_b)/2$, $r'=r_a-r_b$. Once we evaluate the frequency integrals, we then obtain
\begin{equation}\label{FourierTransformsGamma3}
\begin{split}
\Gamma(ij,ab;\Omega)&=\sum_{Q\in BZ}e^{iQ(R-R')}\left(-\frac{2i}{N}\sum_{q\in BZ}\frac{\cos(qr)\cos(qr')}{\Omega-S(q,Q)}\right)\\
&=\sum_{Q\in BZ}e^{iQ(R-R')}\Gamma(r,r';Q,\Omega),
\end{split}
\end{equation}
where $S(q,Q)$ is the two free magnon dispersion, defined in Eq. \eqref{TwoMagnonDispersion} in the main text. Similarly, we can Fourier transform and rewrite $K(lm,pq;t)$ as
\begin{equation}\label{FourierTransformsK3}
\begin{split}
&K(ab,pv;\Omega)=\sum_{Q\in BZ}e^{iQ(R'-R'')}\sum_{q\in BZ}\frac{2i}{N}\frac{\cos(qr'')}{\Omega-S(q,Q)}\cross \\
&\left(8iU(r')\cos(qr')-2i\left(V(r')-i\frac{\Gamma(r')}{2}\right)\cos(Qr'/2)\right)\\
&=\sum_{Q\in BZ}e^{iQ(R'-R'')}K(r',r'';Q,\Omega).
\end{split}
\end{equation}
where $R''=(r_p+r_v)/2$ and $r''=r_p-r_v$. Transforming Eq. \eqref{NewMatrixEquation} by inserting the results of Eq. \eqref{FourierTransformsGamma3} and Eq. \eqref{FourierTransformsK3} gives
\begin{equation}\label{SolnGF2EOM}
\begin{split}
&\frac{1}{N}\int_{-\infty}^{\infty}dt\sum_{Q\in BZ}e^{iQ(R-R')}e^{-i\Omega t}\biggr[G(r,r',Q,\Omega)-\\
&\Gamma(r,r',Q,\Omega)h(r)+\sum_{r''}^{N}K(r,r'',Q,\Omega)G(r',r'',Q,\Omega)\biggr]=0.
\end{split}
\end{equation}
This equation is obeyed provided we set the integrand to zero such that
\begin{equation}\label{App-FTTwoMagnonEOM}
\begin{split}
&\sum_{r''}^{N}\biggr[\delta(r',r'')+K(r',r'',Q,\Omega)\biggr]G(r,r'',Q,\Omega)\\
&=\Gamma(r,r',Q,\Omega)h(r).
\end{split}
\end{equation}
The bound state solutions are found when the determinant of the matrix $\delta(r',r'')+K(r',r'',Q,\Omega)$ is singular, which means $G(r,r'',Q,\Omega)$ cannot be written as the sum of two free magnon solutions. The bound state solutions are therefore solutions to
\begin{equation}\label{App-DetEq}
\begin{split}
&\det\biggr[\delta(r',r'')-\frac{2}{N}\sum_{q\in BZ}8U(r')\frac{\cos(qr')\cos(qr'')}{\Omega-S(q,Q)}+ \\
&\frac{2}{N}\sum_{q\in BZ}\left[2V(r')-i\Gamma(r')\right]\frac{\cos(Qr'/2)\cos(qr'')}{\Omega-S(q,Q)}\biggr]=0.
\end{split}
\end{equation}
If the Ising interaction is nearest-neighbour such that $U_{il}=J_z\delta_{l,i+1}$, we can simplify the determinant in Eq. \eqref{App-DetEq} to obtain Eq. \eqref{NewBoundStates} in the main text.

\section{Simplifying the Determinant Condition}
We first define the Ising and XY matrices,
\begin{equation}\label{}
\begin{split}
&ZZ_{rr'}=\frac{8J_z}{N}\sum_{q\in BZ}\frac{\cos(qr)\cos(qr')}{\Omega-S(q,Q)}\\
&XY_{rr'}=B_rA_{r'},\\
\end{split}
\end{equation}
where
\begin{equation}\label{}
\begin{split}
&A_{r'}=\frac{4}{N}\sum_{q\in BZ}\frac{\cos(qr')}{\Omega-S(q,Q)}\\
&B_r=\left(V(r)-i\frac{\Gamma(r)}{2}\right)\cos(Qr/2).
\end{split}
\end{equation}
This allows us to rewrite the determinant condition, Eq. \eqref{App-DetEq}, as
\begin{equation}\label{}
\begin{split}
&\det(\boldsymbol{I}+\boldsymbol{ZZ}+\boldsymbol{XY}) \\
&=\left|\begin{pmatrix}B_1A_{1}+ZZ_{11}+1& B_2A_{1}  & . & . & B_NA_{1} \\ B_1A_{2}+ZZ_{21} & B_2A_{2}+1   & . & . & B_NA_{2}\\ . & . & . & . & . \\  .  & . & . & . & .\\
B_1A_{N}+ZZ_{N1} & B_2A_{N}  & . & . & B_NA_{N}+1\end{pmatrix}\right|\\
&=\left|\begin{pmatrix}A_{1}+\frac{ZZ_{11}}{B_1}+\frac{1}{B_1}& A_{1} & . & . & A_{N} \\ A_{2}+\frac{ZZ_{21}}{B_2} & A_{2}+\frac{1}{B_2}   & . & . & A_{N} \\  .  & . & . & . & . \\  . & . & . & . & .\\
A_{N}+\frac{ZZ_{N1}}{B_N} & A_{N} & . & . & A_{N}+\frac{1}{B_N}\end{pmatrix}\right|B_1..B_N.\\
\end{split}
\end{equation}
The determinant can be simplified by subtracting the last column from all the other columns, $\text{C}_1-\text{C}_N$, $\text{C}_2-\text{C}_N$,...$\text{C}_{N-1}-\text{C}_N$, giving
\begin{equation}\label{}
\begin{split}
&(ZZ_{11}+1)\left|\begin{pmatrix}  1  & 0 & . & . & 0 & B_2A_{2}\\  0  & 1 & . & . & . & . \\   .  & . & . & . & . & . \\  .  & . & . & . & 0 & . \\   0  & . & . & 0 & 1 & .\\ -\frac{B_1}{B_N}  & . & . & . & . & B_NA_{N}+1\end{pmatrix}\right|\\
&+(-1)^N(B_1A_{1})\left|\begin{pmatrix}  ZZ_{21}  & 1 & 0 & . &. & 0\\  ZZ_{31}  & 0 & 1 & . &. & . \\ . & . &. & . & . & . \\ . &. & . & . & . & 0 \\   ZZ_{(N-1)1} & 0 &. & . & 0 & 1\\ ZZ_{N1}-\frac{B_1}{B_N}  & -\frac{B_2}{B_N} &. & . & . & -B_N+\frac{1}{B_N}\end{pmatrix}\right|,
\end{split}
\end{equation}
where we partially Laplace expand the determinant. For the first determinant, we can swap the first and last column, $C_1\leftrightarrow C_N$ and then swap the first and last row, $R_1\leftrightarrow R_N$. In the second determinant, we can carry out the row-swap operation, $R_N\leftrightarrow R_{N-1}$, followed by $R_{N-1}\leftrightarrow R_{N-2}$, $R_{N-2}\leftrightarrow R_{N-3}$ etc. until the last row becomes the first row. This then gives
\begin{equation}\label{laststep}
\begin{split}
&(ZZ_{11}+1)\left|\begin{pmatrix}  B_N A_{N}+1  & -\frac{B_3}{B_N} & . & .& -\frac{B_{N-1}}{B_N} & -\frac{B_2}{B_N}\\ B_3 A_{3}  & 1 & 0 & . & . & 0 \\   . & 0 & 1 & . & . & . \\   . & .  & . & . & . & . \\ B_{N-1}A_{N-1} & . & . & . & . & . \\ B_{2}A_{2} & 0 & . & . & 0 & 1\end{pmatrix}\right|\\
\\
&-(B_1A_{1})\left|\begin{pmatrix}  ZZ_{N1}-\frac{B_1}{B_N}  & -\frac{B_2}{B_N} & -\frac{B_3}{B_N} & . & . & \frac{B_{N-1}}{B_N}\\  ZZ_{21}  & 1 & 0 & . & . & 0 \\   .  & 0 & 1 & . & . &. \\   .  & . & . & . & . &. \\ ZZ_{(N-2)1}  & . & . & . & . & 0 \\ ZZ_{(N-1)1}  & 0 & . & . & 0 & 1\end{pmatrix}\right|.\\
\end{split}
\end{equation}
which are the determinants of arrowhead matrices, where an arrowhead matrix is a matrix of the form
\begin{equation}\label{arrowhead}
\begin{split}
\boldsymbol{G}=\begin{pmatrix} a & b_2  & b_3 & . & . & b_N \\ c_2 & d_2  & 0 & . & . & 0\\ c_3 & 0  & d_3 & . & . & . \\  . & .  & . & . & . & . \\ . & .  & . & . & . & . \\ c_N & 0  & . & . & . & d_N\end{pmatrix}.\\
\end{split}
\end{equation}
Using the Sherman-Morrison-Woodbury formula, we can evaluate the determinant of the arrowhead matrix by rewriting Eq. \eqref{arrowhead} as
\begin{equation}\label{}
\begin{split}
\det(\boldsymbol{G})&=det(\boldsymbol{A}+\boldsymbol{C}\boldsymbol{B}^T)\\
&=\det(\boldsymbol{I}+\boldsymbol{B}^T\boldsymbol{A}^{(-1)}\boldsymbol{C})det(\boldsymbol{A}),
\end{split}
\end{equation}
where 
\begin{equation}\label{}
\begin{split}&\boldsymbol{A}=
\begin{pmatrix} a & 0 & 0 & . & . & 0 \\ 0 & d_2  & 0 & . & . & 0\\ 0 & 0  & . & . & . & . \\  . & .  & . & . & . & . \\
0 & 0  & . & . & . & d_N\end{pmatrix}\\
&\boldsymbol{B}=
\begin{pmatrix} b_1 & b_2  & b_3 & . & . & b_N \\ 1 & 0  & 0 & . & . & 0\end{pmatrix}\\
&\boldsymbol{C}^T=
\begin{pmatrix} c_1 & c_2  & c_3 & . & . & c_N \\ 1 & 0  & 0 & . & . & 0\end{pmatrix}.
\end{split}
\end{equation}
Using this gives a determinant of
\small
\begin{equation}\label{}
\begin{split}
\det(\boldsymbol{G})=\left[a-\sum_{i=1}^{N}\frac{b_ic_i}{d_i}\right]\prod_{2}^{N} d_i.
\end{split}
\end{equation}
\normalsize
Substituting the values of $a$, $b_i$, $c_i$ and $d_i$ for the two arrowhead matrices in Eq. \eqref{laststep}, we obtain the determinant equation
\begin{equation}\label{finaldet}
\begin{split}
\det(\boldsymbol{G})=(ZZ_{11}+1)(1+\tr(\boldsymbol{XY}))-A_{11}\sum_{i=1}^{N}ZZ_{i1}B_i.
\end{split}
\end{equation}
Once we plug in the definitions of $\boldsymbol{ZZ}$ and $\boldsymbol{XY}$ into Eq. \eqref{finaldet}, we obtain Eq. \eqref{NewBoundStates} in the main text. 

\section{Nearest-Neighbour Bound State Solution}
Here we derive the analytic expression for the bound state energy and decay rate given by Eq. \eqref{NN Solution} when the XY interaction and nonlocal dissipation is nearest-neighbour. We can evaluate the integrals as defined in Eq. \eqref{Integrals} using contour integration. Substituting $z=\exp(iq)$, the integral transforms into
\begin{equation}\label{APP-Integral-NN}
\begin{split}
I_m(t,Q)=\frac{-1}{2^{m}}\oint\frac{(z+z^{-1})^m}{\alpha z^2-(\Omega+t)z+\alpha}\frac{dz}{2\pi i},
\end{split}
\end{equation}
where we have defined $\alpha=(2V_{12}+i\Gamma_{12})\cos(Qa/2)$ and $t=8J_z-2\Delta+i\Gamma$. The integral has a pole of order $m$ at $z=0$ and simple poles at $z_{\pm}=(\Omega+t)/2\alpha\pm\sqrt{((\Omega+t)/2\alpha)^2-1}$. The two poles only coincide at $|z|=1$, so the case of double poles can be ignored for the derivation. Evaluating the integrals gives
\begin{equation}\label{APP-IntegralsEvaluated-NN}
\begin{split}
&I_0(t,Q)=-\frac{\pm 1}{\sqrt{(\Omega+t)^2-4\alpha^2}}\\
&I_1(t,Q)=-\frac{1}{\alpha}-\frac{(\Omega+t)}{2\alpha}\frac{\pm 1}{\sqrt{(\Omega+t)^2-4\alpha^2}}\\
&I_2(t,Q)=-\frac{(\Omega+t)}{\alpha^2}-\frac{(\Omega+t)^2}{4\alpha^2}\frac{\pm 1}{\sqrt{(\Omega+t)^2-4\alpha^2}},
\end{split}
\end{equation}
where the $\pm 1$ sign depends on whether $z_+$ or $z_-$ lie in the contour. Substituting these solutions into the bound state equation, Eq. \eqref{Continuum}, we obtain the equation
\begin{equation}\label{APP-NNBsCond}
\begin{split}
\frac{\pm 1}{\sqrt{(\Omega+t)^2-\alpha^2}}\left(2J_z\frac{(\Omega+t)}{\alpha^2}-1\right)+\frac{2J_z}{\alpha^2}=0,
\end{split}
\end{equation}
which gives the solution $\Omega+t=4+\alpha^2/(4J_z)$. 

\section{Next-Nearest-Neighbour Bound State Solution}
To derive the analytic expression for the next-nearest-neighbour bound state solution given by Eq. \eqref{NNNBSFull}, we use the substitution $z=e^{iq}$ to transform the integral in Eq. \eqref{Integrals} into the following contour integral
\begin{equation}\label{APP-Integral-NNN}
\begin{split}
I_m(t,Q)=\frac{-1}{2^{m}}\oint\frac{z(z+z^{-1})^m}{\beta z^4+\alpha z^3-(\Omega+t) z^2+\alpha z+ \beta}\frac{dz}{2\pi i},
\end{split}
\end{equation}
where $\beta=(2V_{13}-i\Gamma_{13})\cos(Qa)$, $t=8J_z-2\Delta+i\Gamma$ and $\alpha=(2V_{12}-i\Gamma_{12})\cos(Qa/2)$. The quartic in the denominator is palindromic, which means the solutions obey a quadratic in $(z+1/z)$. Therefore, if $z$ is a solution to the quartic, then so too is $1/z$, and this immediately indicates that only two of the four roots can exist inside the contour. We also find that the residue of the roots $1/z$ and $z$ only differ by a sign. The integrals in Eq. \eqref{APP-Integral-NNN} can therefore be evaluated to give
\begin{equation}\label{APP-IntegralsEvaluated-NNN}
\begin{split}
&I_0(t,Q)=\frac{-1}{\beta}(F_1+F_2)\\
&I_1(t,Q)=\frac{-1}{2\beta}(\beta_1F_1+\beta_2F_2)\\
&I_2(t,Q)=\frac{-1}{4\beta}(1+\beta_1^2F_1+\beta_2^2F_2),
\end{split}
\end{equation}
where
\begin{equation}\label{APP-F}
\begin{split}
&F_{1/2}=\pm\frac{1}{\sqrt{\beta_{1/2}^2-4}(\beta_{1/2}-\beta_{2/1})}\\
&\beta_{1/2}=\frac{\alpha}{2\beta}\mp\sqrt{\left(\frac{\alpha}{2\beta}\right)^2+\frac{(\Omega+t)}{\beta}+2}.
\end{split}
\end{equation}
The sign of $F_{1/2}$ depends on whether the root $z_{1/2}$ or its inverse lies inside the contour. Substituting the integral solutions into the bound state equation, Eq. \eqref{Continuum}, gives
\begin{equation}\label{APP-BSCond-NNN}
\begin{split}
\frac{1}{F_1}+\frac{1}{F_2}+\frac{2J_z(\beta_1-\beta_2)^2}{\beta+2J_z}=0.
\end{split}
\end{equation}
We can now solve Eq. \eqref{APP-BSCond-NNN} to obtain the solution given in Eq. \eqref{NNNBSFull} in the main text. There is also the possibility of a double root when $\Omega+t=2\beta+\alpha^2/(4\beta^2)$. When this is the case, the denominator the integrals in Eq. \eqref{Continuum} can be simplified to $(4\beta \cos(q)-\alpha)^2/(4\beta)$. We can then evaluate the NNN integrals without using contour integration, but find these solutions do not obey the bound state solution.
\end{appendices}

\nocite{*}

\begin{thebibliography}{39}%
	\makeatletter
	\providecommand \@ifxundefined [1]{%
		\@ifx{#1\undefined}
	}%
	\providecommand \@ifnum [1]{%
		\ifnum #1\expandafter \@firstoftwo
		\else \expandafter \@secondoftwo
		\fi
	}%
	\providecommand \@ifx [1]{%
		\ifx #1\expandafter \@firstoftwo
		\else \expandafter \@secondoftwo
		\fi
	}%
	\providecommand \natexlab [1]{#1}%
	\providecommand \enquote  [1]{``#1''}%
	\providecommand \bibnamefont  [1]{#1}%
	\providecommand \bibfnamefont [1]{#1}%
	\providecommand \citenamefont [1]{#1}%
	\providecommand \href@noop [0]{\@secondoftwo}%
	\providecommand \href [0]{\begingroup \@sanitize@url \@href}%
	\providecommand \@href[1]{\@@startlink{#1}\@@href}%
	\providecommand \@@href[1]{\endgroup#1\@@endlink}%
	\providecommand \@sanitize@url [0]{\catcode `\\12\catcode `\$12\catcode
		`\&12\catcode `\#12\catcode `\^12\catcode `\_12\catcode `\%12\relax}%
	\providecommand \@@startlink[1]{}%
	\providecommand \@@endlink[0]{}%
	\providecommand \url  [0]{\begingroup\@sanitize@url \@url }%
	\providecommand \@url [1]{\endgroup\@href {#1}{\urlprefix }}%
	\providecommand \urlprefix  [0]{URL }%
	\providecommand \Eprint [0]{\href }%
	\providecommand \doibase [0]{http://dx.doi.org/}%
	\providecommand \selectlanguage [0]{\@gobble}%
	\providecommand \bibinfo  [0]{\@secondoftwo}%
	\providecommand \bibfield  [0]{\@secondoftwo}%
	\providecommand \translation [1]{[#1]}%
	\providecommand \BibitemOpen [0]{}%
	\providecommand \bibitemStop [0]{}%
	\providecommand \bibitemNoStop [0]{.\EOS\space}%
	\providecommand \EOS [0]{\spacefactor3000\relax}%
	\providecommand \BibitemShut  [1]{\csname bibitem#1\endcsname}%
	\let\auto@bib@innerbib\@empty
	\bibitem [{\citenamefont {Yan}\ \emph {et~al.}(2013)\citenamefont {Yan},
		\citenamefont {Moses}, \citenamefont {Gadway}, \citenamefont {Covey},
		\citenamefont {Hazzard}, \citenamefont {Rey}, \citenamefont {Jin},\ and\
		\citenamefont {Ye}}]{Yan2013}%
	\BibitemOpen
	\bibfield  {author} {\bibinfo {author} {\bibfnamefont {B.}~\bibnamefont
			{Yan}}, \bibinfo {author} {\bibfnamefont {S.~A.}\ \bibnamefont {Moses}},
		\bibinfo {author} {\bibfnamefont {B.}~\bibnamefont {Gadway}}, \bibinfo
		{author} {\bibfnamefont {J.~P.}\ \bibnamefont {Covey}}, \bibinfo {author}
		{\bibfnamefont {K.~R.~A.}\ \bibnamefont {Hazzard}}, \bibinfo {author}
		{\bibfnamefont {A.~M.}\ \bibnamefont {Rey}}, \bibinfo {author} {\bibfnamefont
			{D.~S.}\ \bibnamefont {Jin}}, \ and\ \bibinfo {author} {\bibfnamefont
			{J.}~\bibnamefont {Ye}},\ }\href {\doibase 10.1038/nature12483} {\bibfield
		{journal} {\bibinfo  {journal} {Nature}\ }\textbf {\bibinfo {volume} {501}},\
		\bibinfo {pages} {521} (\bibinfo {year} {2013})}\BibitemShut {NoStop}%
	\bibitem [{\citenamefont {Labuhn}\ \emph {et~al.}(2016)\citenamefont {Labuhn},
		\citenamefont {Barredo}, \citenamefont {Ravets}, \citenamefont
		{de~L{\'{e}}s{\'{e}}leuc}, \citenamefont {Macr{\`{i}}}, \citenamefont
		{Lahaye},\ and\ \citenamefont {Browaeys}}]{Labuhn2016}%
	\BibitemOpen
	\bibfield  {author} {\bibinfo {author} {\bibfnamefont {H.}~\bibnamefont
			{Labuhn}}, \bibinfo {author} {\bibfnamefont {D.}~\bibnamefont {Barredo}},
		\bibinfo {author} {\bibfnamefont {S.}~\bibnamefont {Ravets}}, \bibinfo
		{author} {\bibfnamefont {S.}~\bibnamefont {de~L{\'{e}}s{\'{e}}leuc}},
		\bibinfo {author} {\bibfnamefont {T.}~\bibnamefont {Macr{\`{i}}}}, \bibinfo
		{author} {\bibfnamefont {T.}~\bibnamefont {Lahaye}}, \ and\ \bibinfo {author}
		{\bibfnamefont {A.}~\bibnamefont {Browaeys}},\ }\href {\doibase
		10.1038/nature18274} {\bibfield  {journal} {\bibinfo  {journal} {Nature}\
		}\textbf {\bibinfo {volume} {534}},\ \bibinfo {pages} {667} (\bibinfo {year}
		{2016})}\BibitemShut {NoStop}%
	\bibitem [{\citenamefont {Fukuhara}\ \emph {et~al.}(2013)\citenamefont
		{Fukuhara}, \citenamefont {Schau{\ss}}, \citenamefont {Endres}, \citenamefont
		{Hild}, \citenamefont {Cheneau}, \citenamefont {Bloch},\ and\ \citenamefont
		{Gross}}]{Fukuhara2013}%
	\BibitemOpen
	\bibfield  {author} {\bibinfo {author} {\bibfnamefont {T.}~\bibnamefont
			{Fukuhara}}, \bibinfo {author} {\bibfnamefont {P.}~\bibnamefont
			{Schau{\ss}}}, \bibinfo {author} {\bibfnamefont {M.}~\bibnamefont {Endres}},
		\bibinfo {author} {\bibfnamefont {S.}~\bibnamefont {Hild}}, \bibinfo {author}
		{\bibfnamefont {M.}~\bibnamefont {Cheneau}}, \bibinfo {author} {\bibfnamefont
			{I.}~\bibnamefont {Bloch}}, \ and\ \bibinfo {author} {\bibfnamefont
			{C.}~\bibnamefont {Gross}},\ }\href {\doibase 10.1038/nature12541} {\bibfield
		{journal} {\bibinfo  {journal} {Nature}\ }\textbf {\bibinfo {volume}
			{502}},\ \bibinfo {pages} {76} (\bibinfo {year} {2013})}\BibitemShut
	{NoStop}%
	\bibitem [{\citenamefont {Bethe}(1931)}]{Bethe1931}%
	\BibitemOpen
	\bibfield  {author} {\bibinfo {author} {\bibfnamefont {H.}~\bibnamefont
			{Bethe}},\ }\href {\doibase 10.1007/BF01341708} {\bibfield  {journal}
		{\bibinfo  {journal} {Zeitschrift f�r Phys.}\ }\textbf {\bibinfo {volume}
			{71}},\ \bibinfo {pages} {205} (\bibinfo {year} {1931})}\BibitemShut
	{NoStop}%
	\bibitem [{\citenamefont {Wortis}(1963)}]{Wortis1963}%
	\BibitemOpen
	\bibfield  {author} {\bibinfo {author} {\bibfnamefont {M.}~\bibnamefont
			{Wortis}},\ }\href {\doibase 10.1103/PhysRev.132.85} {\bibfield  {journal}
		{\bibinfo  {journal} {Phys. Rev.}\ }\textbf {\bibinfo {volume} {132}},\
		\bibinfo {pages} {85} (\bibinfo {year} {1963})}\BibitemShut {NoStop}%
	\bibitem [{\citenamefont {Haldane}(1982)}]{Haldane1982}%
	\BibitemOpen
	\bibfield  {author} {\bibinfo {author} {\bibfnamefont {F.~D.~M.}\
			\bibnamefont {Haldane}},\ }\href {\doibase 10.1088/0022-3719/15/36/008}
	{\bibfield  {journal} {\bibinfo  {journal} {J. Phys. C Solid State Phys.}\
		}\textbf {\bibinfo {volume} {15}},\ \bibinfo {pages} {L1309} (\bibinfo {year}
		{1982})}\BibitemShut {NoStop}%
	\bibitem [{\citenamefont {Southern}\ \emph {et~al.}(1994)\citenamefont
		{Southern}, \citenamefont {Lee},\ and\ \citenamefont {Lavis}}]{Southern1994}%
	\BibitemOpen
	\bibfield  {author} {\bibinfo {author} {\bibfnamefont {B.~W.}\ \bibnamefont
			{Southern}}, \bibinfo {author} {\bibfnamefont {R.~J.}\ \bibnamefont {Lee}}, \
		and\ \bibinfo {author} {\bibfnamefont {D.~A.}\ \bibnamefont {Lavis}},\ }\href
	{\doibase 10.1088/0953-8984/6/46/024} {\bibfield  {journal} {\bibinfo
			{journal} {J. Phys. Condens. Matter}\ }\textbf {\bibinfo {volume} {6}},\
		\bibinfo {pages} {10075} (\bibinfo {year} {1994})}\BibitemShut {NoStop}%
	\bibitem [{\citenamefont {Schneider}(1981)}]{Schneider1981}%
	\BibitemOpen
	\bibfield  {author} {\bibinfo {author} {\bibfnamefont {T.}~\bibnamefont
			{Schneider}},\ }\href {\doibase 10.1103/PhysRevB.24.5327} {\bibfield
		{journal} {\bibinfo  {journal} {Phys. Rev. B}\ }\textbf {\bibinfo {volume}
			{24}},\ \bibinfo {pages} {5327} (\bibinfo {year} {1981})}\BibitemShut
	{NoStop}%
	\bibitem [{\citenamefont {Torrance}\ and\ \citenamefont
		{Tinkham}(1969)}]{Torrance1969}%
	\BibitemOpen
	\bibfield  {author} {\bibinfo {author} {\bibfnamefont {J.~B.}\ \bibnamefont
			{Torrance}}\ and\ \bibinfo {author} {\bibfnamefont {M.}~\bibnamefont
			{Tinkham}},\ }\href {\doibase 10.1103/PhysRev.187.587} {\bibfield  {journal}
		{\bibinfo  {journal} {Phys. Rev.}\ }\textbf {\bibinfo {volume} {187}},\
		\bibinfo {pages} {587} (\bibinfo {year} {1969})}\BibitemShut {NoStop}%
	\bibitem [{\citenamefont {Majumdar}(1969)}]{Majumdar1969}%
	\BibitemOpen
	\bibfield  {author} {\bibinfo {author} {\bibfnamefont {C.~K.}\ \bibnamefont
			{Majumdar}},\ }\href {\doibase 10.1063/1.1664749} {\bibfield  {journal}
		{\bibinfo  {journal} {J. Math. Phys.}\ }\textbf {\bibinfo {volume} {10}},\
		\bibinfo {pages} {177} (\bibinfo {year} {1969})}\BibitemShut {NoStop}%
	\bibitem [{\citenamefont {Ono}\ \emph {et~al.}(1971)\citenamefont {Ono},
		\citenamefont {Mikado},\ and\ \citenamefont {Oguchi}}]{Ono1971}%
	\BibitemOpen
	\bibfield  {author} {\bibinfo {author} {\bibfnamefont {I.}~\bibnamefont
			{Ono}}, \bibinfo {author} {\bibfnamefont {S.}~\bibnamefont {Mikado}}, \ and\
		\bibinfo {author} {\bibfnamefont {T.}~\bibnamefont {Oguchi}},\ }\href
	{\doibase 10.1143/JPSJ.30.358} {\bibfield  {journal} {\bibinfo  {journal} {J.
				Phys. Soc. Japan}\ }\textbf {\bibinfo {volume} {30}},\ \bibinfo {pages} {358}
		(\bibinfo {year} {1971})}\BibitemShut {NoStop}%
	\bibitem [{\citenamefont {Letscher}\ and\ \citenamefont
		{Petrosyan}(2018)}]{Letscher2018a}%
	\BibitemOpen
	\bibfield  {author} {\bibinfo {author} {\bibfnamefont {F.}~\bibnamefont
			{Letscher}}\ and\ \bibinfo {author} {\bibfnamefont {D.}~\bibnamefont
			{Petrosyan}},\ }\href {\doibase 10.1103/PhysRevA.97.043415} {\bibfield
		{journal} {\bibinfo  {journal} {Phys. Rev. A}\ }\textbf {\bibinfo {volume}
			{97}},\ \bibinfo {pages} {043415} (\bibinfo {year} {2018})}\BibitemShut
	{NoStop}%
	\bibitem [{\citenamefont {Kecke}\ \emph {et~al.}(2007)\citenamefont {Kecke},
		\citenamefont {Momoi},\ and\ \citenamefont {Furusaki}}]{Kecke2007}%
	\BibitemOpen
	\bibfield  {author} {\bibinfo {author} {\bibfnamefont {L.}~\bibnamefont
			{Kecke}}, \bibinfo {author} {\bibfnamefont {T.}~\bibnamefont {Momoi}}, \ and\
		\bibinfo {author} {\bibfnamefont {A.}~\bibnamefont {Furusaki}},\ }\href
	{\doibase 10.1103/PhysRevB.76.060407} {\bibfield  {journal} {\bibinfo
			{journal} {Phys. Rev. B}\ }\textbf {\bibinfo {volume} {76}},\ \bibinfo
		{pages} {060407} (\bibinfo {year} {2007})}\BibitemShut {NoStop}%
	\bibitem [{\citenamefont {Qin}\ \emph {et~al.}(2017)\citenamefont {Qin},
		\citenamefont {Mei}, \citenamefont {Ke}, \citenamefont {Zhang},\ and\
		\citenamefont {Lee}}]{Qin2017}%
	\BibitemOpen
	\bibfield  {author} {\bibinfo {author} {\bibfnamefont {X.}~\bibnamefont
			{Qin}}, \bibinfo {author} {\bibfnamefont {F.}~\bibnamefont {Mei}}, \bibinfo
		{author} {\bibfnamefont {Y.}~\bibnamefont {Ke}}, \bibinfo {author}
		{\bibfnamefont {L.}~\bibnamefont {Zhang}}, \ and\ \bibinfo {author}
		{\bibfnamefont {C.}~\bibnamefont {Lee}},\ }\href {\doibase
		10.1103/PhysRevB.96.195134} {\bibfield  {journal} {\bibinfo  {journal} {Phys.
				Rev. B}\ }\textbf {\bibinfo {volume} {96}},\ \bibinfo {pages} {195134}
		(\bibinfo {year} {2017})}\BibitemShut {NoStop}%
	\bibitem [{\citenamefont {Qin}\ \emph {et~al.}(2018)\citenamefont {Qin},
		\citenamefont {Mei}, \citenamefont {Ke}, \citenamefont {Zhang},\ and\
		\citenamefont {Lee}}]{Qin2018}%
	\BibitemOpen
	\bibfield  {author} {\bibinfo {author} {\bibfnamefont {X.}~\bibnamefont
			{Qin}}, \bibinfo {author} {\bibfnamefont {F.}~\bibnamefont {Mei}}, \bibinfo
		{author} {\bibfnamefont {Y.}~\bibnamefont {Ke}}, \bibinfo {author}
		{\bibfnamefont {L.}~\bibnamefont {Zhang}}, \ and\ \bibinfo {author}
		{\bibfnamefont {C.}~\bibnamefont {Lee}},\ }\href {\doibase
		10.1088/1367-2630/aa9556} {\bibfield  {journal} {\bibinfo  {journal} {New J.
				Phys.}\ }\textbf {\bibinfo {volume} {20}},\ \bibinfo {pages} {013003}
		(\bibinfo {year} {2018})}\BibitemShut {NoStop}%
	\bibitem [{\citenamefont {Agarwala}\ and\ \citenamefont
		{Sen}(2017)}]{Agarwala2017}%
	\BibitemOpen
	\bibfield  {author} {\bibinfo {author} {\bibfnamefont {A.}~\bibnamefont
			{Agarwala}}\ and\ \bibinfo {author} {\bibfnamefont {D.}~\bibnamefont {Sen}},\
	}\href {\doibase 10.1103/PhysRevB.96.104309} {\bibfield  {journal} {\bibinfo
		{journal} {Phys. Rev. B}\ }\textbf {\bibinfo {volume} {96}},\ \bibinfo
	{pages} {104309} (\bibinfo {year} {2017})}\BibitemShut {NoStop}%
\bibitem [{\citenamefont {Kudo}\ \emph {et~al.}(2009)\citenamefont {Kudo},
	\citenamefont {Boness},\ and\ \citenamefont {Monteiro}}]{Kudo2009}%
\BibitemOpen
\bibfield  {author} {\bibinfo {author} {\bibfnamefont {K.}~\bibnamefont
		{Kudo}}, \bibinfo {author} {\bibfnamefont {T.}~\bibnamefont {Boness}}, \ and\
	\bibinfo {author} {\bibfnamefont {T.~S.}\ \bibnamefont {Monteiro}},\ }\href
{\doibase 10.1103/PhysRevA.80.063409} {\bibfield  {journal} {\bibinfo
		{journal} {Phys. Rev. A}\ }\textbf {\bibinfo {volume} {80}},\ \bibinfo
	{pages} {063409} (\bibinfo {year} {2009})}\BibitemShut {NoStop}%
\bibitem [{\citenamefont {Barker}\ \emph {et~al.}(2013)\citenamefont {Barker},
	\citenamefont {Atxitia}, \citenamefont {Ostler}, \citenamefont {Hovorka},
	\citenamefont {Chubykalo-Fesenko},\ and\ \citenamefont
	{Chantrell}}]{Barker2013}%
\BibitemOpen
\bibfield  {author} {\bibinfo {author} {\bibfnamefont {J.}~\bibnamefont
		{Barker}}, \bibinfo {author} {\bibfnamefont {U.}~\bibnamefont {Atxitia}},
	\bibinfo {author} {\bibfnamefont {T.~A.}\ \bibnamefont {Ostler}}, \bibinfo
	{author} {\bibfnamefont {O.}~\bibnamefont {Hovorka}}, \bibinfo {author}
	{\bibfnamefont {O.}~\bibnamefont {Chubykalo-Fesenko}}, \ and\ \bibinfo
	{author} {\bibfnamefont {R.~W.}\ \bibnamefont {Chantrell}},\ }\href {\doibase
	10.1038/srep03262} {\bibfield  {journal} {\bibinfo  {journal} {Sci. Rep.}\
	}\textbf {\bibinfo {volume} {3}},\ \bibinfo {pages} {3262} (\bibinfo {year}
	{2013})}\BibitemShut {NoStop}%
\bibitem [{\citenamefont {Krimphoff}\ \emph {et~al.}(2017)\citenamefont
	{Krimphoff}, \citenamefont {Haque},\ and\ \citenamefont
	{L{\"{a}}uchli}}]{Krimphoff2017}%
\BibitemOpen
\bibfield  {author} {\bibinfo {author} {\bibfnamefont {C.~B.}\ \bibnamefont
		{Krimphoff}}, \bibinfo {author} {\bibfnamefont {M.}~\bibnamefont {Haque}}, \
	and\ \bibinfo {author} {\bibfnamefont {A.~M.}\ \bibnamefont
		{L{\"{a}}uchli}},\ }\href {\doibase 10.1103/PhysRevB.95.144308} {\bibfield
	{journal} {\bibinfo  {journal} {Phys. Rev. B}\ }\textbf {\bibinfo {volume}
		{95}},\ \bibinfo {pages} {144308} (\bibinfo {year} {2017})}\BibitemShut
{NoStop}%
\bibitem [{\citenamefont {Ganahl}\ \emph {et~al.}(2012)\citenamefont {Ganahl},
	\citenamefont {Rabel}, \citenamefont {Essler},\ and\ \citenamefont
	{Evertz}}]{Gana}%
\BibitemOpen
\bibfield  {author} {\bibinfo {author} {\bibfnamefont {M.}~\bibnamefont
		{Ganahl}}, \bibinfo {author} {\bibfnamefont {E.}~\bibnamefont {Rabel}},
	\bibinfo {author} {\bibfnamefont {F.~H.~L.}\ \bibnamefont {Essler}}, \ and\
	\bibinfo {author} {\bibfnamefont {H.~G.}\ \bibnamefont {Evertz}},\ }\href
{\doibase 10.1103/PhysRevLett.108.077206} {\bibfield  {journal} {\bibinfo
		{journal} {Phys. Rev. Lett.}\ }\textbf {\bibinfo {volume} {108}},\ \bibinfo
	{pages} {077206} (\bibinfo {year} {2012})}\BibitemShut {NoStop}%
\bibitem [{\citenamefont {M{\"{o}}lter}\ \emph {et~al.}(2014)\citenamefont
	{M{\"{o}}lter}, \citenamefont {Barthel}, \citenamefont {Schollw{\"{o}}ck},\
	and\ \citenamefont {Alba}}]{Molter2014}%
\BibitemOpen
\bibfield  {author} {\bibinfo {author} {\bibfnamefont {J.}~\bibnamefont
		{M{\"{o}}lter}}, \bibinfo {author} {\bibfnamefont {T.}~\bibnamefont
		{Barthel}}, \bibinfo {author} {\bibfnamefont {U.}~\bibnamefont
		{Schollw{\"{o}}ck}}, \ and\ \bibinfo {author} {\bibfnamefont
		{V.}~\bibnamefont {Alba}},\ }\href {\doibase
	10.1088/1742-5468/2014/10/P10029} {\bibfield  {journal} {\bibinfo  {journal}
		{J. Stat. Mech. Theory Exp.}\ }\textbf {\bibinfo {volume} {2014}},\ \bibinfo
	{pages} {P10029} (\bibinfo {year} {2014})}\BibitemShut {NoStop}%
\bibitem [{\citenamefont {Longo}\ and\ \citenamefont
	{Evers}(2014{\natexlab{a}})}]{Longo2014}%
\BibitemOpen
\bibfield  {author} {\bibinfo {author} {\bibfnamefont {P.}~\bibnamefont
		{Longo}}\ and\ \bibinfo {author} {\bibfnamefont {J.}~\bibnamefont {Evers}},\
}\href {\doibase 10.1103/PhysRevLett.112.193601} {\bibfield  {journal}
{\bibinfo  {journal} {Phys. Rev. Lett.}\ }\textbf {\bibinfo {volume} {112}},\
\bibinfo {pages} {193601} (\bibinfo {year} {2014}{\natexlab{a}})}\BibitemShut
{NoStop}%
\bibitem [{\citenamefont {Hsu}\ \emph {et~al.}(2016)\citenamefont {Hsu},
	\citenamefont {Zhen}, \citenamefont {Stone}, \citenamefont {Joannopoulos},\
	and\ \citenamefont {Solja{\v{c}}i{\'{c}}}}]{Hsu2016}%
\BibitemOpen
\bibfield  {author} {\bibinfo {author} {\bibfnamefont {C.~W.}\ \bibnamefont
		{Hsu}}, \bibinfo {author} {\bibfnamefont {B.}~\bibnamefont {Zhen}}, \bibinfo
	{author} {\bibfnamefont {A.~D.}\ \bibnamefont {Stone}}, \bibinfo {author}
	{\bibfnamefont {J.~D.}\ \bibnamefont {Joannopoulos}}, \ and\ \bibinfo
	{author} {\bibfnamefont {M.}~\bibnamefont {Solja{\v{c}}i{\'{c}}}},\ }\href
{\doibase 10.1038/natrevmats.2016.48} {\bibfield  {journal} {\bibinfo
		{journal} {Nat. Rev. Mater.}\ }\textbf {\bibinfo {volume} {1}},\ \bibinfo
	{pages} {16048} (\bibinfo {year} {2016})}\BibitemShut {NoStop}%
\bibitem [{\citenamefont {Longo}\ and\ \citenamefont
	{Evers}(2014{\natexlab{b}})}]{Longo2014c}%
\BibitemOpen
\bibfield  {author} {\bibinfo {author} {\bibfnamefont {P.}~\bibnamefont
		{Longo}}\ and\ \bibinfo {author} {\bibfnamefont {J.}~\bibnamefont {Evers}},\
}\href {\doibase 10.1103/PhysRevA.90.063834} {\bibfield  {journal} {\bibinfo
	{journal} {Phys. Rev. A}\ }\textbf {\bibinfo {volume} {90}},\ \bibinfo
{pages} {063834} (\bibinfo {year} {2014}{\natexlab{b}})}\BibitemShut
{NoStop}%
\bibitem [{\citenamefont {Schauss}(2018)}]{Schauss2018}%
\BibitemOpen
\bibfield  {author} {\bibinfo {author} {\bibfnamefont {P.}~\bibnamefont
		{Schauss}},\ }\href {\doibase 10.1088/2058-9565/aa9c59} {\bibfield  {journal}
	{\bibinfo  {journal} {Quantum Sci. Technol.}\ }\textbf {\bibinfo {volume}
		{3}},\ \bibinfo {pages} {023001} (\bibinfo {year} {2018})}\BibitemShut
{NoStop}%
\bibitem [{\citenamefont {Weimer}\ \emph {et~al.}(2010)\citenamefont {Weimer},
	\citenamefont {M{\"{u}}ller}, \citenamefont {Lesanovsky}, \citenamefont
	{Zoller},\ and\ \citenamefont {B{\"{u}}chler}}]{Weimer2010}%
\BibitemOpen
\bibfield  {author} {\bibinfo {author} {\bibfnamefont {H.}~\bibnamefont
		{Weimer}}, \bibinfo {author} {\bibfnamefont {M.}~\bibnamefont
		{M{\"{u}}ller}}, \bibinfo {author} {\bibfnamefont {I.}~\bibnamefont
		{Lesanovsky}}, \bibinfo {author} {\bibfnamefont {P.}~\bibnamefont {Zoller}},
	\ and\ \bibinfo {author} {\bibfnamefont {H.~P.}\ \bibnamefont
		{B{\"{u}}chler}},\ }\href {\doibase 10.1038/nphys1614} {\bibfield  {journal}
	{\bibinfo  {journal} {Nat. Phys.}\ }\textbf {\bibinfo {volume} {6}},\
	\bibinfo {pages} {382} (\bibinfo {year} {2010})}\BibitemShut {NoStop}%
\bibitem [{\citenamefont {Whitlock}\ \emph {et~al.}(2017)\citenamefont
	{Whitlock}, \citenamefont {Glaetzle},\ and\ \citenamefont
	{Hannaford}}]{Whitlock2016}%
\BibitemOpen
\bibfield  {author} {\bibinfo {author} {\bibfnamefont {S.}~\bibnamefont
		{Whitlock}}, \bibinfo {author} {\bibfnamefont {A.~W.}\ \bibnamefont
		{Glaetzle}}, \ and\ \bibinfo {author} {\bibfnamefont {P.}~\bibnamefont
		{Hannaford}},\ }\href {\doibase 10.1088/1361-6455/aa6149} {\bibfield
	{journal} {\bibinfo  {journal} {J. Phys. B At. Mol. Opt. Phys.}\ }\textbf
	{\bibinfo {volume} {50}},\ \bibinfo {pages} {074001} (\bibinfo {year}
	{2017})}\BibitemShut {NoStop}%
\bibitem [{\citenamefont {Nguyen}\ \emph {et~al.}(2018)\citenamefont {Nguyen},
	\citenamefont {Raimond}, \citenamefont {Sayrin}, \citenamefont
	{Corti{\~{n}}as}, \citenamefont {Cantat-Moltrecht}, \citenamefont {Assemat},
	\citenamefont {Dotsenko}, \citenamefont {Gleyzes}, \citenamefont {Haroche},
	\citenamefont {Roux}, \citenamefont {Jolicoeur},\ and\ \citenamefont
	{Brune}}]{Nguyen2018}%
\BibitemOpen
\bibfield  {author} {\bibinfo {author} {\bibfnamefont {T.~L.}\ \bibnamefont
		{Nguyen}}, \bibinfo {author} {\bibfnamefont {J.~M.}\ \bibnamefont {Raimond}},
	\bibinfo {author} {\bibfnamefont {C.}~\bibnamefont {Sayrin}}, \bibinfo
	{author} {\bibfnamefont {R.}~\bibnamefont {Corti{\~{n}}as}}, \bibinfo
	{author} {\bibfnamefont {T.}~\bibnamefont {Cantat-Moltrecht}}, \bibinfo
	{author} {\bibfnamefont {F.}~\bibnamefont {Assemat}}, \bibinfo {author}
	{\bibfnamefont {I.}~\bibnamefont {Dotsenko}}, \bibinfo {author}
	{\bibfnamefont {S.}~\bibnamefont {Gleyzes}}, \bibinfo {author} {\bibfnamefont
		{S.}~\bibnamefont {Haroche}}, \bibinfo {author} {\bibfnamefont
		{G.}~\bibnamefont {Roux}}, \bibinfo {author} {\bibfnamefont {T.}~\bibnamefont
		{Jolicoeur}}, \ and\ \bibinfo {author} {\bibfnamefont {M.}~\bibnamefont
		{Brune}},\ }\href {\doibase 10.1103/PhysRevX.8.011032} {\bibfield  {journal}
	{\bibinfo  {journal} {Phys. Rev. X}\ }\textbf {\bibinfo {volume} {8}},\
	\bibinfo {pages} {011032} (\bibinfo {year} {2018})}\BibitemShut {NoStop}%
\bibitem [{\citenamefont {Glaetzle}\ \emph {et~al.}(2015)\citenamefont
	{Glaetzle}, \citenamefont {Dalmonte}, \citenamefont {Nath}, \citenamefont
	{Gross}, \citenamefont {Bloch},\ and\ \citenamefont {Zoller}}]{Glaetzle2015}%
\BibitemOpen
\bibfield  {author} {\bibinfo {author} {\bibfnamefont {A.~W.}\ \bibnamefont
		{Glaetzle}}, \bibinfo {author} {\bibfnamefont {M.}~\bibnamefont {Dalmonte}},
	\bibinfo {author} {\bibfnamefont {R.}~\bibnamefont {Nath}}, \bibinfo {author}
	{\bibfnamefont {C.}~\bibnamefont {Gross}}, \bibinfo {author} {\bibfnamefont
		{I.}~\bibnamefont {Bloch}}, \ and\ \bibinfo {author} {\bibfnamefont
		{P.}~\bibnamefont {Zoller}},\ }\href {\doibase
	10.1103/PhysRevLett.114.173002} {\bibfield  {journal} {\bibinfo  {journal}
		{Phys. Rev. Lett.}\ }\textbf {\bibinfo {volume} {114}},\ \bibinfo {pages}
	{173002} (\bibinfo {year} {2015})}\BibitemShut {NoStop}%
\bibitem [{\citenamefont {Hartmann}(2016)}]{Hartmann2016}%
\BibitemOpen
\bibfield  {author} {\bibinfo {author} {\bibfnamefont {M.~J.}\ \bibnamefont
		{Hartmann}},\ }\href {\doibase 10.1088/2040-8978/18/10/104005} {\bibfield
	{journal} {\bibinfo  {journal} {J. Opt.}\ }\textbf {\bibinfo {volume} {18}},\
	\bibinfo {pages} {104005} (\bibinfo {year} {2016})}\BibitemShut {NoStop}%
\bibitem [{\citenamefont {Hood}\ \emph {et~al.}(2016)\citenamefont {Hood},
	\citenamefont {Goban}, \citenamefont {Asenjo-Garcia}, \citenamefont {Lu},
	\citenamefont {Yu}, \citenamefont {Chang},\ and\ \citenamefont
	{Kimble}}]{Hood2016}%
\BibitemOpen
\bibfield  {author} {\bibinfo {author} {\bibfnamefont {J.~D.}\ \bibnamefont
		{Hood}}, \bibinfo {author} {\bibfnamefont {A.}~\bibnamefont {Goban}},
	\bibinfo {author} {\bibfnamefont {A.}~\bibnamefont {Asenjo-Garcia}}, \bibinfo
	{author} {\bibfnamefont {M.}~\bibnamefont {Lu}}, \bibinfo {author}
	{\bibfnamefont {S.-P.}\ \bibnamefont {Yu}}, \bibinfo {author} {\bibfnamefont
		{D.~E.}\ \bibnamefont {Chang}}, \ and\ \bibinfo {author} {\bibfnamefont
		{H.~J.}\ \bibnamefont {Kimble}},\ }\href {\doibase 10.1073/pnas.1603788113}
{\bibfield  {journal} {\bibinfo  {journal} {Proc. Natl. Acad. Sci.}\ }\textbf
	{\bibinfo {volume} {113}},\ \bibinfo {pages} {10507} (\bibinfo {year}
	{2016})}\BibitemShut {NoStop}%
\bibitem [{\citenamefont {Goban}\ \emph {et~al.}(2014)\citenamefont {Goban},
	\citenamefont {Hung}, \citenamefont {Yu}, \citenamefont {Hood}, \citenamefont
	{Muniz}, \citenamefont {Lee}, \citenamefont {Martin}, \citenamefont
	{McClung}, \citenamefont {Choi}, \citenamefont {Chang}, \citenamefont
	{Painter},\ and\ \citenamefont {Kimble}}]{Goban2014}%
\BibitemOpen
\bibfield  {author} {\bibinfo {author} {\bibfnamefont {A.}~\bibnamefont
		{Goban}}, \bibinfo {author} {\bibfnamefont {C.-L.}\ \bibnamefont {Hung}},
	\bibinfo {author} {\bibfnamefont {S.-P.}\ \bibnamefont {Yu}}, \bibinfo
	{author} {\bibfnamefont {J.}~\bibnamefont {Hood}}, \bibinfo {author}
	{\bibfnamefont {J.}~\bibnamefont {Muniz}}, \bibinfo {author} {\bibfnamefont
		{J.}~\bibnamefont {Lee}}, \bibinfo {author} {\bibfnamefont {M.}~\bibnamefont
		{Martin}}, \bibinfo {author} {\bibfnamefont {A.}~\bibnamefont {McClung}},
	\bibinfo {author} {\bibfnamefont {K.}~\bibnamefont {Choi}}, \bibinfo {author}
	{\bibfnamefont {D.}~\bibnamefont {Chang}}, \bibinfo {author} {\bibfnamefont
		{O.}~\bibnamefont {Painter}}, \ and\ \bibinfo {author} {\bibfnamefont
		{H.}~\bibnamefont {Kimble}},\ }\href {\doibase 10.1038/ncomms4808} {\bibfield
	{journal} {\bibinfo  {journal} {Nat. Commun.}\ }\textbf {\bibinfo {volume}
		{5}},\ \bibinfo {pages} {3808} (\bibinfo {year} {2014})}\BibitemShut
{NoStop}%
\bibitem [{\citenamefont {Douglas}\ \emph {et~al.}(2015)\citenamefont
	{Douglas}, \citenamefont {Habibian}, \citenamefont {Hung}, \citenamefont
	{Gorshkov}, \citenamefont {Kimble},\ and\ \citenamefont
	{Chang}}]{Douglas2015}%
\BibitemOpen
\bibfield  {author} {\bibinfo {author} {\bibfnamefont {J.~S.}\ \bibnamefont
		{Douglas}}, \bibinfo {author} {\bibfnamefont {H.}~\bibnamefont {Habibian}},
	\bibinfo {author} {\bibfnamefont {C.-L.}\ \bibnamefont {Hung}}, \bibinfo
	{author} {\bibfnamefont {A.~V.}\ \bibnamefont {Gorshkov}}, \bibinfo {author}
	{\bibfnamefont {H.~J.}\ \bibnamefont {Kimble}}, \ and\ \bibinfo {author}
	{\bibfnamefont {D.~E.}\ \bibnamefont {Chang}},\ }\href {\doibase
	10.1038/nphoton.2015.57} {\bibfield  {journal} {\bibinfo  {journal} {Nat.
			Photonics}\ }\textbf {\bibinfo {volume} {9}},\ \bibinfo {pages} {326}
	(\bibinfo {year} {2015})}\BibitemShut {NoStop}%
\bibitem [{\citenamefont {Gonz{\'{a}}lez-Tudela}\ \emph
	{et~al.}(2015)\citenamefont {Gonz{\'{a}}lez-Tudela}, \citenamefont {Hung},
	\citenamefont {Chang}, \citenamefont {Cirac},\ and\ \citenamefont
	{Kimble}}]{Gonz}%
\BibitemOpen
\bibfield  {author} {\bibinfo {author} {\bibfnamefont {A.}~\bibnamefont
		{Gonz{\'{a}}lez-Tudela}}, \bibinfo {author} {\bibfnamefont {C.-L.}\
		\bibnamefont {Hung}}, \bibinfo {author} {\bibfnamefont {D.~E.}\ \bibnamefont
		{Chang}}, \bibinfo {author} {\bibfnamefont {J.~I.}\ \bibnamefont {Cirac}}, \
	and\ \bibinfo {author} {\bibfnamefont {H.~J.}\ \bibnamefont {Kimble}},\
}\href {\doibase 10.1038/nphoton.2015.54} {\bibfield  {journal} {\bibinfo
	{journal} {Nat. Photonics}\ }\textbf {\bibinfo {volume} {9}},\ \bibinfo
{pages} {320} (\bibinfo {year} {2015})}\BibitemShut {NoStop}%
\bibitem [{\citenamefont {Calaj{\'{o}}}\ \emph {et~al.}(2016)\citenamefont
	{Calaj{\'{o}}}, \citenamefont {Ciccarello}, \citenamefont {Chang},\ and\
	\citenamefont {Rabl}}]{Calajo2016}%
\BibitemOpen
\bibfield  {author} {\bibinfo {author} {\bibfnamefont {G.}~\bibnamefont
		{Calaj{\'{o}}}}, \bibinfo {author} {\bibfnamefont {F.}~\bibnamefont
		{Ciccarello}}, \bibinfo {author} {\bibfnamefont {D.}~\bibnamefont {Chang}}, \
	and\ \bibinfo {author} {\bibfnamefont {P.}~\bibnamefont {Rabl}},\ }\href
{\doibase 10.1103/PhysRevA.93.033833} {\bibfield  {journal} {\bibinfo
		{journal} {Phys. Rev. A}\ }\textbf {\bibinfo {volume} {93}},\ \bibinfo
	{pages} {033833} (\bibinfo {year} {2016})}\BibitemShut {NoStop}%
\bibitem [{\citenamefont {Zeiher}\ \emph {et~al.}(2016)\citenamefont {Zeiher},
	\citenamefont {van Bijnen}, \citenamefont {Schau{\ss}}, \citenamefont {Hild},
	\citenamefont {Choi}, \citenamefont {Pohl}, \citenamefont {Bloch},\ and\
	\citenamefont {Gross}}]{Zeiher2016}%
\BibitemOpen
\bibfield  {author} {\bibinfo {author} {\bibfnamefont {J.}~\bibnamefont
		{Zeiher}}, \bibinfo {author} {\bibfnamefont {R.}~\bibnamefont {van Bijnen}},
	\bibinfo {author} {\bibfnamefont {P.}~\bibnamefont {Schau{\ss}}}, \bibinfo
	{author} {\bibfnamefont {S.}~\bibnamefont {Hild}}, \bibinfo {author}
	{\bibfnamefont {J.-y.}\ \bibnamefont {Choi}}, \bibinfo {author}
	{\bibfnamefont {T.}~\bibnamefont {Pohl}}, \bibinfo {author} {\bibfnamefont
		{I.}~\bibnamefont {Bloch}}, \ and\ \bibinfo {author} {\bibfnamefont
		{C.}~\bibnamefont {Gross}},\ }\href {\doibase 10.1038/nphys3835} {\bibfield
	{journal} {\bibinfo  {journal} {Nat. Phys.}\ }\textbf {\bibinfo {volume}
		{12}},\ \bibinfo {pages} {1095} (\bibinfo {year} {2016})}\BibitemShut
{NoStop}%
\bibitem [{\citenamefont {Perrella}\ \emph {et~al.}(2018)\citenamefont
	{Perrella}, \citenamefont {Light}, \citenamefont {Vahid}, \citenamefont
	{Benabid},\ and\ \citenamefont {Luiten}}]{Perrella2018}%
\BibitemOpen
\bibfield  {author} {\bibinfo {author} {\bibfnamefont {C.}~\bibnamefont
		{Perrella}}, \bibinfo {author} {\bibfnamefont {P.~S.}\ \bibnamefont {Light}},
	\bibinfo {author} {\bibfnamefont {S.~A.}\ \bibnamefont {Vahid}}, \bibinfo
	{author} {\bibfnamefont {F.}~\bibnamefont {Benabid}}, \ and\ \bibinfo
	{author} {\bibfnamefont {A.~N.}\ \bibnamefont {Luiten}},\ }\href {\doibase
	10.1103/PhysRevApplied.9.044001} {\bibfield  {journal} {\bibinfo  {journal}
		{Phys. Rev. Appl.}\ }\textbf {\bibinfo {volume} {9}},\ \bibinfo {pages}
	{044001} (\bibinfo {year} {2018})}\BibitemShut {NoStop}%
\bibitem [{\citenamefont {Breuer}\ and\ \citenamefont
	{Petruccione}(2007)}]{Breuer2007}%
\BibitemOpen
\bibfield  {author} {\bibinfo {author} {\bibfnamefont {H.-P.}\ \bibnamefont
		{Breuer}}\ and\ \bibinfo {author} {\bibfnamefont {F.}~\bibnamefont
		{Petruccione}},\ }\href {\doibase 10.1093/acprof:oso/9780199213900.001.0001}
{\emph {\bibinfo {title} {{The Theory of Open Quantum Systems}}}}\ (\bibinfo
{publisher} {Oxford University Press},\ \bibinfo {year} {2007})\BibitemShut
{NoStop}%
\bibitem [{\citenamefont {Majlis}(2000)}]{Majlis2000}%
\BibitemOpen
\bibfield  {author} {\bibinfo {author} {\bibfnamefont {N.}~\bibnamefont
		{Majlis}},\ }\href {\doibase 10.1142/4178} {\emph {\bibinfo {title} {{The
				Quantum Theory of Magnetism}}}}\ (\bibinfo  {publisher} {World Scientific},\
\bibinfo {year} {2000})\BibitemShut {NoStop}%
\end{thebibliography}

%

\end{document}